\documentclass{jfm}
\usepackage{array}   
\usepackage{graphicx}
\usepackage{dcolumn}
\usepackage{lscape}
\usepackage{bm}
\usepackage[utf8]{inputenc}
\usepackage{subfig}
\usepackage{graphicx}
\usepackage{ragged2e}
\usepackage{amsmath}
\usepackage{mathtools}
\usepackage{caption}
\usepackage{color}
\usepackage[most]{tcolorbox}
\usepackage{xcolor}
\usepackage{float}
\usepackage{bigints}
\usepackage[export]{adjustbox}
\usepackage{tabularx}
\usepackage{epstopdf, epsfig}
\usepackage[toc,page]{appendix}
\usepackage{nomencl}
\usepackage{mathtools,amssymb,lipsum}
\usepackage{url}
\usepackage{newtxtext}
\usepackage{newtxmath}
\usepackage{natbib}
\usepackage{hyperref}
\usepackage{mathtools}
\usepackage{booktabs}   
\usepackage[normalem]{ulem}
\usepackage{makecell}
\hypersetup{
	colorlinks = true,
	urlcolor   = pink,
	citecolor  = blue,
}

\usepackage{booktabs}   
\usepackage{tabularx}  
\usepackage{caption}   
\usepackage{xcolor}    
\usepackage{xcolor}

\usepackage{multirow}
\usepackage{xcolor}
\usepackage{colortbl}

\usepackage{siunitx}
\usepackage{soul}

\newcommand{\RomanNumeralCaps}[1]
\linenumbers

\newcommand{\vp}{\phi}

\def\XXint#1#2#3{{\setbox0=\hbox{$#1{#2#3}{\int}$}
		\vcenter{\hbox{$#2#3$}}\kern-.5\wd0}}

\definecolor{huntergreen}{HTML}{355E3B}

\title{Interfacial waves from pressure forcing: \textcolor{magenta}{role of the transients}}
\author{Vinod Kumar Kadari\aff{1}, Nikhil Yewale\aff{1}, Palas Kumar Farsoiya\aff{2}, Y. S. Mayya\aff{1}\corresp{\email{ysmayya@gmail.com}} \and Ratul Dasgupta\aff{1}\corresp{\email{dasgupta.ratul@gmail.com}}}

\affiliation{\aff{1}Chemical Engineering, Indian Institute of Technology Bombay, India 400 076,\aff{2}Chemical Engineering, Indian Institute of Technology Roorkee, India }

\begin{document}
	\maketitle
	\newcommand{\mj}{{\mathrm{J}}}
	\begin{abstract}
		 A localised overpressure translating at an uniform speed $U > U_c$, acts at the interface separating two deep fluid layers with densities $\rho_u < \rho_l$ (upper and lower). We analyse \textcolor{magenta}{this problem from the perspective of eliciting the role of transients in} obtaining solutions for the wave patterns of this configuration. \textcolor{magenta}{This is done} through an initial-value problem (IVP) formulation, within the linearised, inviscid, potential flow approximation. It is shown that the steady-state interface (as time $\tilde{t}\rightarrow\infty$) featuring short (capillary) and long (gravity) waves leading and trailing the overpressure respectively, originate from a peculiar asymmetric cancellation of Fourier components in the far-field of the forcing. The time-dependent part of the solution, relaxing algebraically as $\tilde{t}^{-1/2}$ \textcolor{magenta}{(pure gravity case)} plays a crucial role in this cancellation. This is in marked contrast to the classical approaches \citep{rayleigh1883form, Kelvin1905} of directly solving the steady-state equations of motion, which admit infinitely many solutions, unless one invokes conditions \textit{``ab extra''} such as the radiation boundary condition, artificial dissipation or group-velocity arguments. We generalise the IVP approach of a one-fluid free surface originally proposed by \cite{stoker1953unsteady}, to a two-fluid interface. \textcolor{magenta}{Further, it is shown that the time-dependent Green function to the problem can be used for determining the interface response due to random-in-time, localised pressure forcing (Gaussian white noise). This elegantly recovers the classical result of \cite{phillips1957generation} within our IVP framework. 
		 For a steady point force,} the time evolution of these waves, at low and moderately high density ratios, are also benchmarked against nonlinear simulations - excellent agreement is demonstrated at small forcing strength; for larger forcing, no steady-state is observed. Qualitative similarities between capillary-gravity Stokes/\textcolor{magenta}{Crapper} wave profiles and those seen in our simulations are emphasized. \textcolor{magenta}{The study illustrates the power of the IVP approach towards obtaining insights into the time-dependent behavior of interfacial waves. These results have immediate applications in the study of wind-driven, transient waves as well as in computations of transient, wave-induced drag.}
	\end{abstract}


	\section{Introduction}\label{sec:intro}
\begin{figure}
     	\centering
     	\subfloat[Rayleigh and Kelvin's cavity]{\includegraphics[scale=0.75]{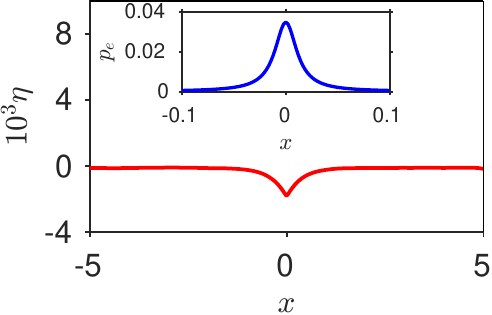}\label{fig1a}}\quad
     	\subfloat[Deep-layer dispersion relation]{\includegraphics[scale=0.75]{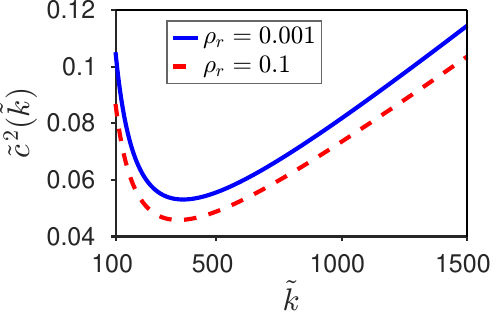}\label{fig1b}}
     	\captionsetup{justification=raggedright,singlelinecheck=false}
     	\caption{Panel (a) Nonlinear simulation of the occurence of a stationary interfacial cavity in a stagnant pool of deep water due to the application of a localised, steady ($\tilde{t}>0$), pressure disturbance of the Lorentzian form (inset) $\tilde{p}_e(\tilde{x}) = \dfrac{\tilde{F}_0}{\pi} \left(\dfrac{\tilde{b}}{\tilde{b}^2 + \tilde{x}^2}\right)$ (see section \ref{sec:nonlin_sim} for implementation details). A subset of the total simulation domain excluding the transient waves is presented. Panel (b) (Dimensional) dispersion relation (S.I. units) for unforced, linearised capillary-gravity waves at the interface of two deep layers i.e. $\tilde{c}^2(\tilde{k}) = \left(\dfrac{\rho_l-\rho_u}{\rho_l + \rho_u}\right)g\tilde{k}^{-1} + \dfrac{T\tilde{k}}{\rho_l + \rho_u}$ for density ratio $\rho_r \equiv \dfrac{\rho_u}{\rho_l} = \{10^{-3},10^{-1}\}$, $g=9.81$ m/s$^2$ and $T=0.072$ N/m.}
     	\label{fig1}
     \end{figure}

     Imagine an experiment where a small blower (e.g. a hair dryer) is switched on and held stationary in  a vertically downward configuration, above the interface of an otherwise stagnant pool of deep water (fig. $1$ in \cite{ghabache2014liquid}). After sufficiently long time, the transient waves at the air-water interface beneath the blower disperse and we intuitively expect the interface to deform into a cavity at steady-state. The shape of this is determined by competition between the applied pressure (air from the blower) which deforms the interface and the combined restoring forces of surface-tension and gravity, which resist this deformation. We refer the interested reader to fig. $25$ in \cite{Kelvin1905} for a two-dimensional presentation of this cavity profile taking into account only gravity; fig. \ref{fig1a} depicts a simulational realization of Rayleigh's version of this cavity considering surface-tension as well as gravity (see above eqn. $14$ in \cite{rayleigh1883form}). \textcolor{magenta}{ In fig. \ref{fig1}, variables with (without) tilde on top are dimensional (non-dimensional). $\tilde{F}_0$ represents force per unit length, $\tilde{b}$ is a measure of the width of the Lorentzian and the deformed interface is scaled as $\eta(x,t) \equiv \tilde{\eta}g\tilde{U}^{-2}$. The non-dimensional pressure is $p_e \equiv \dfrac{\tilde{p}_e}{\rho_l \tilde{U}^2}$ with $\dfrac{\tilde{F}_0}{\rho_l\tilde{U}^4g^{-1}} = 0.0014$, the (non-dimensional) distance $x\equiv \tilde{x}g\tilde{U}^{-2}$. Here $\tilde{U}=0.267$ m/s is a choice of scale for non-dimensionalisation and the plot corresponds to time $t \equiv \tilde{t}g\tilde{U}^{-1}=36.7$.} 
     
     If the blower is now made to travel at constant speed $U$, the interface response can be more complex. Dependending on whether $U$ is less than (subcritical), greater than (supercritical) or nearly equal to (transcritical) a critical value $U_c$, the interfacial response as observed in a co-moving frame, can appear in one of the three forms respectively: (a) localised and steady, with appreciable deformation visible only directly under the blower; (b) steady with delocalised far-field wave pattern; (c) unsteady and nearly periodic in time.  \textcolor{magenta}{In the context of fig. \ref{fig1b}, it is also useful to recall that the critical phase-speed $U_c$ for the super-critical regime ($U > U_c$) is given by $\dfrac{\rho_l U_c^4}{gT} = 4\left[\dfrac{1-\rho_r}{\left(1 + \rho_r\right)^2}\right]$.} When the pattern is steady and delocalised (supercritical regime), the waves upstream and downstream are markedly asymmetric in wavelength; in the co-moving frame short waves appear ahead of the blower whereas longer wavelengths manifest around and behind it (e.g. fig. $6$ in \cite{burghelea2001onset}). While these characteristics of the supercritical regime have been known for a long time, at least since the observations in \cite{russel1844report} (see their plate $57$), some of the sub-critical observations are \textcolor{magenta}{more} recent and summarised in a neat set of experiments in \cite{diorio2011resonantly}, see their fig. $4$. 
     
    The earliest theoretical studies of the steady-state supercritical wave patterns were by \cite{rayleigh1883form}, \cite{Kelvin1905}, \cite{lamb1932hydrodynamics} (hereafter RKL) and a related time-dependent study by \cite{havelock1908propagation}, all under linearised approximation.  
  	The theoretical framework of RKL for the occurrence of asymmetric steady-state
  	patterns upstream and downstream included assumptions or conditions external to the
  	equations of motion. To circumvent these, \textcolor{magenta}{some} authors \citep{havelock1908propagation,stoker1953unsteady,puri1970linear} have proposed that a more satisfactory approach is to look at the steady-state solution as the asymptotic time limit of a suitably posed Initial Value Problem (IVP). \textcolor{magenta}{Apart from this, there are other important applications \citep{phillips1957generation,li2022principal} where intrinsically it is necessary to invoke the time-dependent behaviour of the system. The IVP approach thus, is a
  		logical development that
  		calls for a comprehensive investigation.} With this in view, in this study we focus on the IVP approach for the two-dimensional version of the RKL problem, further generalised to the case of an interface between two liquids of arbitrary density ratio, specifically in the super-critical regime ($U > U_c$). The time-dependent version of the RKL problem will be shown to generate important insights for the steady-state solution (see next sub-section). Although the generalization to arbitrary density ratio has been performed for a moving pressure source in three-dimensions, albeit only for the steady-state by \cite{crapper1967ship}, the unsteady, two-dimensional solution is being reported here for the first time. \textcolor{magenta}{Additionally, as mentioned above the time-dependent approach is essential in several problems of physical interest, discussed later on. With these motivations, the aim of our study is to investigate the RKL problem from an IVP as well as Green's function perspective, to elicit explicit functional forms for the time-dependent solutions, restricting ourselves to the super-critical regime sustaining waves in the far-field.} Table \ref{tab:lit_map} presents a literature summary of waves from pressure and other kinds of forcing; we also refer the reader to chapter $4$, figs. $4.1$, $4.2$, $4.3$ and $4.4$ in the textbook by \cite{vanden2010gravity}, which emphasize commonality of wave response arising due to varied forcing mechanisms. Our table \ref{tab:lit_map} was partly motivated by these insightful images in \cite{vanden2010gravity}.
     
     \subsection{Uniqueness of solutions to steady-state equations}\label{subsec:unique}
     Obtaining unique solutions to the linearised, steady-state equations with pressure forcing featuring waves in the far-field, historically has been a source of intriguing mathematical ideas. \cite{rayleigh1883form} and \cite{Kelvin1905} adopted distinct strategies for mitigating the non-uniqueness. \cite{rayleigh1883form} obtained the steady, capillary-gravity wave response from a surface pressure forcing (Dirac delta function) on a stream moving at uniform speed. To obtain a unique answer for the observed short and long waves upstream and downstream in the supercritical regime, he employed (artifical) dissipation in his Fourier integrals. 
     
     \begin{landscape}
     	\begin{center}
     		\setlength{\tabcolsep}{4pt}
     		\renewcommand{\arraystretch}{0.7}
     		\scriptsize
     		\begin{tabular}{|
     				p{0.8cm}|
     				p{2.5cm}|
     				p{2.5cm}|
     				p{2.3cm}|
     				p{2.3cm}|
     				p{2.3cm}|}
     			\hline
     			& & \multicolumn{2}{|c|}{\textbf{Two-dimensional}}
     			& \multicolumn{2}{|c|}{\textbf{Three-dimensional}} \\ \cline{3-6}
     			
     			\textbf{EOM} \rule{0pt}{4ex} & \centering\textbf{Nature of forcing (localised force/source(s))}
     			& \centering$\mathbf{\dfrac{\partial}{\partial t} = 0}$ & \centering$\mathbf{\dfrac{\partial}{\partial t} \neq 0}$
     			& \centering$\mathbf{\dfrac{\partial}{\partial t}=0}$ & $\mathbf{\dfrac{\partial}{\partial t}\neq0}$ \\
     			\hline
     			
     			\multirow{30}{*}{\textbf{\textit{Laplace}}}
     			& \vspace{1mm}\includegraphics[scale=0.75]{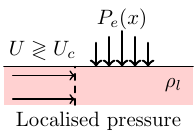}
     			& \vspace{1mm}\textbf{One layer:$\dagger$} \citep{rayleigh1883form,lamb1916lxv,chepelianskii2010self},$(*)$\citep{schwartz1981nonlinear, vanden2002wilton,maleewong2005free_1,maleewong2005free_2},$\dagger$\citep{havelock1922effect} \textbf{\textcolor{magenta}{(long waves)}}, \textbf{Two layers:$\dagger$} \citep{lamb1916xliv}
     			& \vspace{1mm}\textbf{One layer:$\dagger$} \citep{havelock1908propagation,green1948lxxxvii,stoker1953unsteady,wurtele1955transient,puri1970linear},\textbf{Two layers: }\textcolor{orange}{\textbf{Current work}}$(\dagger,*)$
     			& \vspace{1mm}\textbf{One layer:$\dagger$} \citep{ursell1960kelvin,crapper1964surface,raphael1996capillary,shliomis1997surface,burghelea2002wave,reed2002ship,rabaud2013ship}, $\dagger$\citep{cumberbatch1965effects,liang2018asymptotic} \textbf{\textcolor{magenta}{(viscous corrections)}}, \textbf{Two layers:$\dagger$} \citep{crapper1967ship} &
     			\vspace{2mm}\textbf{One layer:$(*)$}\cite{wu1982three} \textbf{\textcolor{magenta}{(long waves)}} \\ \cline{2-6}
     			& \includegraphics[scale=0.75]{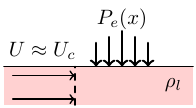}
     			& \vspace{-9mm}\textbf{No steady-state} & \vspace{-9mm}\textbf{One layer:$(*)$} \citep{cho2009forced,cho2011resonantly}, \citep{akylas1984excitation,wu1987generation,ertekin1986waves,casciola1996nonlinear,masnadi2017generation} & \vspace{-9mm}\textbf{No steady-state} & \vspace{-9mm}\textbf{One layer:$(*)$}\citep{akylas1987unsteady}, \cite{katsis1987excitation, wu1982three} \textbf{\textcolor{magenta}{(long waves)}}\\ \cline{2-6}     			
     			&
     			\vspace{1mm}\includegraphics[scale=0.75]{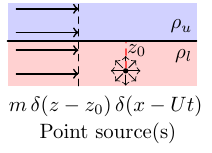}
     			& \vspace{1mm}\textbf{One layer $\dagger$:} \citep{havelock1932theory}, page 489 in \cite{WehausenLaitone}, also \cite{kibel1964theoretical} & \vspace{1mm}\textbf{One layer:$(*)$} \cite{zhu1997nonlinear} \textbf{\textcolor{magenta}{(dipole)}} & \vspace{1mm}\textbf{One layer:} page 483 in $(\dagger)$\citep{WehausenLaitone},  \citep{noblesse1977fundamental,lustri2013steady,liang2019viscous}, \textbf{Two layers:$\dagger$} \citep{hudimac1961ship} & \vspace{1mm}\textbf{One layer:$\dagger$} \cite{lustri2014unsteady}, \textbf{Two layers:$\dagger$} \citep{you1991analytical,yeung1999waves} \\ \cline{1-6}
     			\multirow{4}{0.5cm}{\textit{\textbf{Oseen \& Laplace}}}
     			& \vspace{1mm} \includegraphics[scale=0.75]{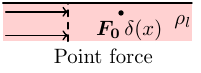}
     			& \vspace{1mm}- & \vspace{1mm}\textbf{One layer:}\citep{lu2008unsteady}$\dagger$ & \vspace{1mm}\textbf{Two layers:$\dagger$} \citep{lu2005interfacial} \textbf{\textcolor{magenta}{(point force and point source)}} & \vspace{1mm}\textbf{One layer:$\dagger$}   \citep{lu2005unsteady} \textbf{\textcolor{magenta}{(viscous layer(s))}} \\
     			\hline
     		\end{tabular}
     		\vspace{0.2em}
     		\captionof{table}{\small Non-exhaustive literature overview classified per equations of motion (EOM), linear ($\dagger$) or nonlinear ($*$), nature of forcing (pressure, point source, point force), spatial dimensionality (2D/3D), number of layers (one/two) and presence/absence of temporal evolution (steady/unsteady).}
     		\label{tab:lit_map}
     	\end{center}
     \end{landscape}

    \cite{Kelvin1905} also examined the interface deformation (neglecting surface tension) for a two-dimensional, localised pressure distribution, a so-called `forcive' of Lorentzian form $\tilde{p}_e(\tilde{x})=\rho g \tilde{a} \left(\dfrac{\tilde{b}^2}{\tilde{x}^2 + \tilde{b}^2}\right)$, where $\rho, g, \tilde{a}$ and $\tilde{b}$ being density, acceleration due to gravity, amplitude and width respectively, of the Lorentzian travelling with speed $U$, see eqn. $95$ in \cite{Kelvin1905} (a different symbol for $\tilde{a}$ was used). Kelvin obtained a \textit{symmetric wave-train}, upstream and downstream of the forcive (see his fig. $29$) but alluded to the "(in)stability" of the wavetrain ahead (behind) of the forcive. These aforementioned observations point at the non-uniqueness of the solutions to the steady-state equations unless conditions \textit{``ab extra''} are imposed. In the theory to follow in section \ref{sec:IVP}, we will present the solution to the \textit{linearised} IVP involving an externally imposed pressure in the form of a Dirac delta function at the interface of two immiscible fluids. It will be seen that the IVP naturally generates an \textit{unique}, asymmetric wave response to the linearised, time-dependent equations without requiring dissipation terms \citep{rayleigh1883form}, invoking instability aguments \citep{Kelvin1905} \textcolor{magenta}{or requiring radiation boundary conditions}. This is to be contrasted against solution to the corresponding steady equations of motion (section \ref{subsec:steady}), which feature arbitrary constants. \textcolor{magenta}{On the one hand, the solution to the IVP naturally determines these constants (see section \ref{subsec:ivp_Fourier} and appendix \ref{AppD-GreenFn}) but furthermore, the transients resolved correctly in the IVP approach have practical relevance to problems of physical interest, as discussed next.}
     \subsection{\textcolor{magenta}{Why are the transients important?}}\label{sec:need}     
     The outstanding example(s) of solution to the \textit{two-dimensional IVP} for pressure-forced waves at finite depth and zero density ratio, are by \cite{stoker1953unsteady} (gravity waves) and \cite{puri1970linear} (capillary-gravity waves). \textcolor{magenta}{In the subsequent literature, the associated radiation condition has been highlighted for resolution of the non-uniqueness of solutions to the steady-state equations \citep{vanden2001damped}}.  
     While uniqueness via the IVP approach was demonstrated by \cite{stoker1953unsteady,puri1970linear} in the $\tilde{t}\rightarrow\infty$ limit, \textcolor{magenta}{systematic analysis of the time-dependent convergence to steady state has not been carried out so far - this gap is addressed here. Importantly, we emphasize that the time-dependent IVP approach has conceptual and practical applicability well-beyond resolution of the aforementioned uniqueness problem. This is explained briefly below via two examples.} 
     	
     \textcolor{magenta}{Consider the problem of generation of surface waves due to travelling, random (in time) pressure forcing (due to turbulent wind fluctuations advected by the mean flow), generating resonant surface waves which travel at the same phase-speed as the pressure forcing (supercritical regime). This was considered originally within an IVP framework (in three dimensions) in a path-breaking study by O. M. \cite{phillips1957generation}. The classical result of \cite{phillips1957generation} shows that in the so-called ``principal-stage of wind-wave development'' (section $4$ in \cite{phillips1957generation}), the (ensemble averaged) mean-squared displacement of the interface grows \textit{linearly with time} (eqn. $4.13$ in \cite{phillips1957generation}) commencing from an initially flat interface. We show in Appendix E that the Phillips result (derived in three-dimensions) is quite elegantly recovered also in two dimensions for a (random-in-time) point force, using the transient Green function within our IVP framework presented in section \ref{sec:IVP}. Furthermore, our analytical results in section \ref{sec:IVP} explicitly resolving the time-dependence for a deterministic, travelling, steady point force, corresponds to the initial stage of wave-development in \cite{phillips1957generation} (section $3$ therein). While the study by \cite{phillips1957generation} was focussed on relating the two-point correlation of interface fluctuations to that of the externally imposed pressure fluctations (see eqn. $3.5$ in \cite{phillips1957generation}), our analysis explicitly resolves the transients for a point steady force (Dirac delta function) and is of potential interest towards understanding the early stage of the wind-wave generation process. Incidentally, \cite{Geva2022} have studied the impulsively started, wind-wave problem and emphasize the importance of transients by writing (begin quote) ``\textit{No theoretical description of the waves’ evolution from initially quiescent water surface to finite steady state is
	currently available}'' (end quote, see their first page, first column, last paragraph). Our work here partly addresses this stated gap, although limited to two-dimensions. This can serve as reference for future three-dimensional calculations.} 

    \textcolor{magenta}{As a second example where transients are important, consider the problem of determining the wave-induced drag on a point force applied on a flowing stream moving with speed $U > U_c$ (supercritical regime). In dimensional variables one can estimate the steady-state drag force $F_{d}^{\left(\infty\right)}$ by balancing the work done by this force per unit time with the energy flux of a (linearised) sinusoidal wave (this appears downstream of the force). For pure gravity waves (no surface tension), this yields $F_{d}^{\left(\infty\right)} = \dfrac{P_0^2g}{\rho_lU^4}$. However, prior to attaining this steady-state value, it is expected that the drag force $F_{d}(t)$ passes through transients given by the time-dependent integral $F_{d}(t) = \int_{-\infty}^{\infty}\tilde{p}_e(\tilde{x},\tilde{t})\dfrac{\partial\tilde{\eta}}{\partial\tilde{x}}(\tilde{x},\tilde{t})d\tilde{x}$, see  e.g. eqn. $2.26$ in \cite{raphael1996capillary}. The transient integrals obtained in this study are of use in computing this time-dependent drag $F_{d}(t)$. This time-dependent, wave-induced drag is of interest in several situations where it features a significant `overshoot' over its steady-state value. See for example, page $4$ in \cite{havelock1949wave} and \cite{Barratt1965} for a practical application.}
    
    \textcolor{magenta}{It is also important to note that the well-known Lighthill-Whitham (LW hereafter) time-dependent approach (described in \cite{lighthill2001waves}, see section $13.9$ in \cite{whitham1974linear}), is a pseudo-transient approach and not the true solution to an IVP. LW use a ``growing exponential in time" pressure forcing (with a small parameter $\epsilon > 0$ in the exponent) starting from $t\rightarrow -\infty$. The process is pseudo-transient because it is introduced a long time ago and as an infinite time has lapsed, all the transients have died down at the current time $t$ except for an exponential pressure multiplier term (eqn. $13.56$ in \cite{whitham1974linear}). On the other hand, a true IVP approach begins at time $t=0$ (or at any other finite time), and retains the transient terms which gradually disappear after the lapse of an infinite time. Indeed `true transients' that feature in a typical IVP solution with well-posed initial conditions, involve Fourier integrals which decay \textit{algebraically} in time (see section \ref{subsubsec:asymp}). The time-dependence in these cannot be factored out of the Fourier integral, as is done in the LW approach.}    
     
     \textcolor{magenta}{We recall that} the supercritical regime of current interest (in the deep layer limit) corresponds to the translation speed $U$ of the pressure disturbance satisfying $U > U_c$. As is well-known \citep{maleewong2005free_1,maleewong2005free_2}, such a minimum phase speed $U_c$ (for \textit{unforced, capillary-gravity} waves of finite wavenumber, at the interface of two immiscible, fluid layers) always exists independent of the density ratio of the layers, provided both are modelled as infinitely deep. It turns out however that on layers of \textit{finite depth}, for the existence of such a minimum in phase speed (at a finite, non-zero wavenumber $\tilde{k}_m$), the Bond number $B$ (see below) must be in the range $0 < B < B_c$ where $B_c$ is a critical value dependent on the density ratio $\left(\rho_r\right)$ and thickness ratio of the layers. For $0 \leq \rho_r < 1$ and with $H \equiv H_u/H_l$ denoting the thickness ratio of the layers, it may be shown that 

     \begin{eqnarray}
     	B_c(\rho_r,H) = \dfrac{1}{3}\dfrac{H\left(1-\rho_r\right)\left(1+H\rho_r\right)}{\left(H+\rho_r\right)}. \label{eqBo-2}
     \end{eqnarray}
     It is verified from eqn. \ref{eqBo-2} that $B_c(\rho_r\rightarrow 0,H)\rightarrow 1/3$ and this generalises the zero density ratio earlier results of \cite{maleewong2005free_1,maleewong2005free_2} to $0 \leq \rho_r < 1$. In section \ref{sec:nonlin_sim}, expression \ref{eqBo-2} has been used to ensure that the Bond number $B$, for all our simulational results verifies $0< B < B_c(\rho_r,H)$, thereby ensuring qualitative aspects of the deep-water limit.
  
     \section{How does the wave pattern develop ?}\label{sec:IVP}
   	\begin{figure}	    	
     	\centering
     	\captionsetup{justification=raggedright,singlelinecheck=false}
     	\includegraphics[scale=0.6]{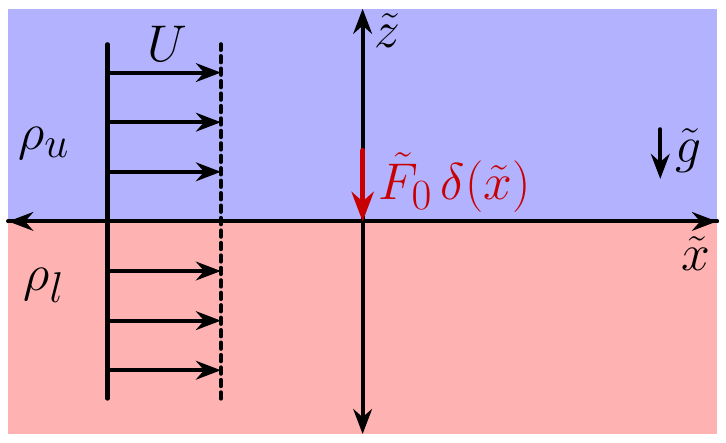}
     	\caption{A point force (red arrow) of strength $\tilde{F}_0$ acts at $\tilde{t}>0$ at the interface of two fluids of density $\rho_u$ and $\rho_l$, both moving with speed $U$ (as seen in the co-moving frame).}
     	\label{fig2}
     \end{figure}
\begin{figure}	    	
      \centering
       \captionsetup{justification=raggedright,singlelinecheck=false}
      \includegraphics[scale=1.1]{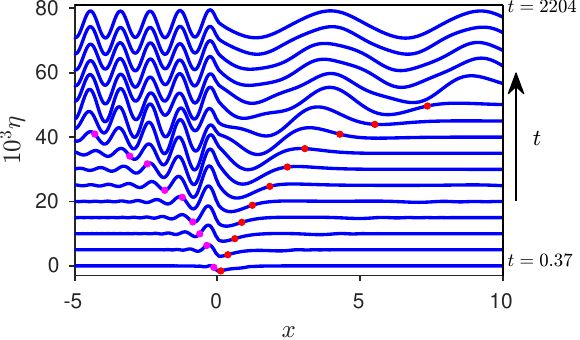}
      \caption{The (non-dimensional) perturbed interface viz. $\eta(x,t) \equiv  \frac{\tilde{\eta}(\tilde{x},\tilde{t})g}{U^2}$ as a function of (non-dimensional) distance ($x$) for various (non-dimensional) time instants ($t$) due to an externally applied Dirac delta function overpressure at $x=0$, as obtained from linearised theory in section \ref{sec:IVP} with $\rho_r=10^{-3}, \alpha=0.1389, F_0=0.0014$; for definition, see below equations \ref{eqIVP-1}(k)-(m). From bottom to top, each interface corresponds to (\textcolor{magenta}{non-dimensional}) time $t=\{0.37, 1.10, 1.84, 2.57, 3.67, 5.51, 7.35, 9.18, 12.86, 16.53, 22.04, 55.10, 110.21, 183.68, 367.35, 2204.1\}$. Each curve is shifted vertically upward except the first one ($t=0.37$). Here $x = \frac{\tilde{x}g}{U^2}$, $t \equiv \dfrac{\tilde{t}g}{U}$. The red and pink dots indicate $\dfrac{x}{t} =c_g^{(-)}|_{k_{s,l}}$ i.e. energy propagation (group) velocity of wavenumbers $k_{s,l}$ (see eqn. \ref{eqIVP-10}) in the co-moving frame (see Appendix \ref{AppB} for $c_g^{(-)}$).}
      \label{fig3}
	  \end{figure}

      We now solve the linearised IVP towards understanding the time evolution of the wave pattern towards steady state, commencing from a flat interface. An external pressure distribution (Dirac delta function at $\tilde{x}=0$) acts at the interface $z=0^{+}$ of two immiscible fluids with density $\rho_u$ and $\rho_l$, surface tension $T$ and gravity $g$, both fluids assumed to be infinitely deep. In the (fixed) laboratory frame of reference, this force moves along the interface leftwards at a speed $U > U_c$. In a frame of reference \textit{co-moving} with this point force, the force appears stationary and both the fluids move with a uniform velocity $U$ rightwards, as indicated in fig. \ref{fig2}. To motivate the theoretical calculation, we anticipate results from our time-dependent linear theory in fig. \ref{fig3}. The figure shows the temporal evolution of the interface leading to the steady-state at $t >> 1$, featuring long (gravity) waves for $x > 0$ and shorter (capillary) ones for $x < 0$. \textcolor{magenta}{For $t = 2204 >> 1$, the steady-state is apparent with notable asymmetry between $x<0$ (short waves) and $x>0$ (long waves). Crucially, \textit{we do not require} radiation boundary conditions at $x\rightarrow\pm \infty$ or dissipation \citep{rayleigh1883form}, to obtain this asymmetric steady-state in the IVP.} Our aim is to develop insight into this time evolution presented in fig. \ref{fig3}; in particular we seek to highlight the origin of the asymmetry about $x=0$ at steady-state ($t >> 1$) and how the time-dependent solution plays an important role in establishing this. 
      
      \subsection{Formulation of governing equations}\label{subsec:ivp_init}
    In this section, we solve the the linearised, inviscid, potential flow equations which also form the cornerstone of subsequent analysis. The justification of these approximations is seen by noting that the time taken to reach the steady-state within a typical observational domain seen in fig. \ref{fig3} is much lower than the time for appreciable vorticity diffusion from the interface into the bulk as well as viscous dissipation; this remains true for air over water or at higher density ratio e.g. water alongwith a mixture of silicone oil and $1$-$2$-$3$-$4$-tetrahydronaphtalene \citep{pouliquen1994propagating}. Owing to the potential flow approximation, the disturbance velocity field may be expressed as the gradient of a velocity potential. The total velocity field can then be written as a sum of a uniform flow (see fig. \ref{fig2}) and a perturbed one  viz. $\mathbf{\tilde{u}}\equiv \left[U,0\right] + \bm{\tilde{\nabla}}\tilde{\vp}$ where $\tilde{\bm{\nabla}}\equiv \left(\dfrac{\partial}{\partial \tilde{x}},\dfrac{\partial}{\partial\tilde{z}}\right)$. The dimensional equations of motion, linearised about the state of uniform velocity $U$ and a flat interface, are provided in Appendix \ref{AppA} Their non-dimensional counterparts (without tildes on top) are:
	\begin{subequations}\label{eqIVP-1}
		\begin{align}
			&\nabla^2\vp_u=0,\quad\quad -\infty < x < \infty,\quad \eta(x,t) \leq z < \infty, \tag{\theequation a}\\			
			& \nabla^2\vp_l=0,\quad\quad -\infty < x < \infty,\quad -\infty < z \leq \eta(x,t), \tag{\theequation b}\\
			& \dfrac{\partial\eta}{\partial t}+ \dfrac{\partial\eta}{\partial x} - \left(\dfrac{\partial\vp_u}{\partial z}\right)_{z=0} = \dfrac{\partial\eta}{\partial t}+ \dfrac{\partial\eta}{\partial x} - \left(\dfrac{\partial\vp_l}{\partial z}\right)_{z=0} =  0, \tag{\theequation c,d} \\
			&-\alpha \left(\dfrac{\partial^2\eta}{\partial x^2}\right) + \Bigg\{\dfrac{\partial \vp_l}{\partial t} -  \rho_r\dfrac{\partial\vp_u}{\partial t} + \left(\dfrac{\partial\vp_l}{\partial x}\right) - \rho_r\left(\dfrac{\partial\vp_u}{\partial x}\right)\bigg\}_{z=0} + \left(1 - \rho_r\right)\eta \nonumber \\
			&= -p_e(x,z=0^{+},t), \tag{\theequation e}\\
			& \bm{\nabla}\phi_l\left(z\rightarrow -\infty\right)\rightarrow \text{finite},\quad \bm{\nabla}\phi_u\left(z\rightarrow \infty\right)\rightarrow \text{finite}, \tag{\theequation f} \\
	    	& \vp_u(x,z=0,t=0)=    		\vp_l(x,z=0,t=0)=0,\;\eta(x,t=0)=0,\tag{\theequation g,h,i,j} \\
	    	& p_e(x,z=0^{+},t> 0) = F_0\delta(x) , \quad F_0 > 0, \tag{\theequation k} \\
	    	& \text{where}\quad \alpha \equiv \dfrac{gT}{\rho_lU^4}, \quad\rho_r \equiv \dfrac{\rho_u}{\rho_l},\quad\beta \equiv \left(\dfrac{1-\rho_r}{1+\rho_r}\right), \tag{\theequation l,m,n} 
    	\end{align}
    \end{subequations}
	$\bm{\nabla} \equiv \left(\dfrac{\partial}{\partial x},\dfrac{\partial}{\partial z}\right)$ and $\nabla^2 \equiv \dfrac{\partial^2}{\partial x^2} + \dfrac{\partial^2}{\partial z^2}$. Eqns. \ref{eqIVP-1}(a, b) are the Laplace eqns. for the velocity potentials, eqns. \ref{eqIVP-1}(c, d) and (e) are the kinematic boundary condition(s) and the Bernoulli equation respectively. Equations \ref{eqIVP-1}(g)-(j) represent initial conditions, while the overpressure term $p_e(x,z=0^{+},t>0)$ acts as a forcing term to eqn. \ref{eqIVP-1}(e) (see form in eqn. \ref{eqIVP-1}(k)). The non-dimensional numbers characterising the system are $\alpha$ \citep{vanden1992gravity}, density ratio $\rho_r$ and the strength of pressure forcing $F_0$, as defined in \ref{eqIVP-1}(k)-(m). In subsequent analysis, the Atwood number $\beta$ defined in \ref{eqIVP-1}(n) occurs frequently and we will use $\beta$ or $\rho_r$, as convenient. In order to obtain explicit  time-dependent solutions to eqns. \ref{eqIVP-1} we proceed as follows.
	
	\subsection{Explicit solutions to eqns. \ref{eqIVP-1} using Fourier transforms}\label{subsec:ivp_Fourier}
	Following \citep{stoker1953unsteady, puri1970linear}, we employ Fourier transforms in the $x$ direction denoted by over bars for the variables in Fourier space. For a typical variable $\Phi(x,z,t)$, its Fourier transform and inverse transform are defined as:
		\begin{subequations}\label{eqIVP-2}
			\begin{align}
					&\bar{\Phi}(k,z,t)\equiv\dfrac{1}{\sqrt{2\pi}}\int_{-\infty}^{\infty}dx\;\exp(-ikx)\Phi(x,z,t),\; \Phi(x,z,t)=\dfrac{1}{\sqrt{2\pi}}\int_{-\infty}^{\infty}dk\;\exp(ikx)\bar{\Phi}(k,z,t), \tag{\theequation a,b}
				\end{align}
			\end{subequations}
	    where $i \equiv \sqrt{-1}$. Our strategy is to convert the first-order (in time) eqns. in \ref{eqIVP-1}, involving three unknowns viz. $\bar{\phi}_u(k;z,t),\bar{\phi}_l(k;z,t)$ and $\bar{\eta}(k;t)$ to a second-order eqn. in time involving only two unknowns. For this, one eliminates $\bar{\eta}(k,t)$ and obtains a second-order eqn. involving $\bar{\vp_{u}}(k,0,t)$ and $\bar{\vp_{l}}(k,0,t)$. After rearrangements (see eqn. $1.9$ of sec. $1.1$ in Supplementary Material), we obtain the following:
	    \begin{eqnarray}\label{eqIVP-3}
			    	&&\left\{\dfrac{\partial^2\bar{\vp}_l}{\partial t^2} + 2ik\dfrac{\partial\bar{\vp}_l}{\partial t} - k^2 \bar{\vp}_l+\left(1+\alpha k^2-\rho_r\right) \dfrac{\partial\bar{\vp}_l}{\partial z}-\rho_r\dfrac{\partial^2\bar{\vp}_u}{\partial t^2}-\rho_r2ik\dfrac{\partial\bar{\vp}_u}{\partial t}+\rho_rk^2 \bar{\vp}_u\right\}_{z=0} \nonumber\\
			    	&&= -ik\bar{p}_e(k),\quad t > 0. 
		\end{eqnarray}
		 where the time-dependence in $\bar{p}_e(k)$ in eqn. \ref{eqIVP-3} has been droppped. The Fourier transformed Laplace equations \ref{eqIVP-1}(a)-(b) are solved subject to boundedness condition at $z\rightarrow\pm\infty$ to obtain
			     \begin{subequations}\label{eqIVP-4}
				     	  \begin{align}	
					     	     \bar{\vp}_l(k;z,t)=A(k,t)\exp\left(|k|z\right),\quad -\infty< z \leq \bar{\eta}(k,t), \nonumber \\
					     	     \bar{\vp}_u(k;z,t)=D(k,t)\exp\left(-|k|z\right),\quad \bar{\eta}(k,t) \leq  z < \infty, \tag{\theequation a,b}
					     	  \end{align} 
				     \end{subequations}
			   Employing the Fourier transformed versions of \ref{eqIVP-1}(c)-(d), and eliminating $D(k,t)$, we obtain the following equation for the unknown coefficient $A(k,t)$ appearing in eqn. \ref{eqIVP-4}(a):
			   \begin{eqnarray}\label{eqIVP-5}
				   		\left(1+\rho_r\right)\frac{\partial^2 A}{\partial t^2}
				   		& +\; 2\left(1+\rho_r\right) i k \dfrac{\partial A}{\partial t}
				   		+ \bigg\{\alpha|k|^3 - \left(1+\rho_r\right)|k|^2 + \left(1-\rho_r\right)|k|\bigg\} A
				   		= - i k\, \bar{p}_e(k), \nonumber \\
				   \end{eqnarray}
			which is to be solved with the initial conditions $A(k,t=0)=\left(\dfrac{\partial A}{\partial t}\right)_{t=0}=0$. The solution to eqn. \ref{eqIVP-5} satisfying the aforementioned initial conditions is (see Supplementary Material sec. $1.1$ for details):
			\begin{eqnarray}\label{eqIVP-6}
				\dfrac{A(k,t)}{\bar{p}_e(k,t)}
				&=& 
				\left(
				\dfrac{ik}{-\alpha|k|^3 + (1+\rho_r)|k|^2 - (1-\rho_r)|k|}
				\right)\\ 
				&+&\dfrac{1}{2}
				\dfrac{ik}{\left(1+\rho_r\right)^{1/2}\big(\alpha|k|^3 + (1-\rho_r)|k|\big)^{1/2}}
				\Bigg(
				\dfrac{\exp\left[-it\lambda_2(k)\right]}{\lambda_2(k)}
				-
				\dfrac{\exp\left[-it\lambda_1(k)\right]}{\lambda_1(k)}
				\Bigg), \nonumber\\
				&\text{where}&\;\lambda_{1,2}(k) \equiv k\mp\sqrt{\dfrac{\alpha|k|^3}{1+\rho_r}\;+\;\beta|k|}\quad .  \nonumber
			\end{eqnarray}
			
			For our choice of the form of the overpressure, $p_e(x,t>0)=F_0\delta(x)$ implies $\bar{p}_e(k,t>0)=\dfrac{F_0}{\sqrt{2\pi}}$. With this and the expression for $A(k,t)$ (eqn. \ref{eqIVP-6}), one obtains (after rearrangements, see  sec. $1.1$ of Supplementary Material) the following expression for the perturbed interface $\bar{\eta}(k,t)$ in the Fourier domain:
			\begin{eqnarray}\label{eqIVP-7}
				\dfrac{\sqrt{2\pi}\;\bar{\eta}(k,t)}{F_0}
				&=&\left(\dfrac{1}{-\alpha|k|^2 + \left(1+\rho_r\right)|k| - \left(1-\rho_r\right)}\right)  \nonumber \\
				&-& \dfrac{1}{\alpha |k|^2 + \left(1 - \rho_r\right)}
				\left(\dfrac{k}{2}\right) \Bigg(
				\dfrac{\exp\left[-it\lambda_2(k)\right]}{\lambda_2(k)}
				+
				\dfrac{\exp\left[-it\lambda_1(k)\right]}{\lambda_1(k)}
				\Bigg), \nonumber \\
			\end{eqnarray}
			The expression for $\eta(x,t)$ may then be obtained from eqn. \ref{eqIVP-7} using the inverse Fourier integral
			\begin{eqnarray}\label{eqIVP-8}
				\eta(x,t) = \dfrac{1}{\sqrt{2\pi}}\int_{-\infty}^{\infty}dk\;\exp\left(ikx\right)\bar{\eta}(k,t), 
			\end{eqnarray}
			and the steady-state solution will emerge from this, in the limit $t\rightarrow\infty$.
			\subsection{Non-uniqueness of the steady-state solutions}
			\label{subsec:steady}
			At first glance, it appears that the time-independent state is contained entirely in the first term on the right hand side of eqn. \ref{eqIVP-7}, in view of the fact that the time-dependent terms in this are expected to vanish as $t\rightarrow\infty$ due to the oscillatory nature of the \textcolor{magenta}{integrands}. The time-independent term has simple poles which lie on the real $k$ axis for our super-critical regime of interest. However, the term is a function only of $|k|$ and hence it is expected that its Fourier inverse in real space will be symmetric about $x=0$, \textcolor{magenta}{implying symmetric response upstream and downstream of the forcing - this is contrary to observation}.
			On the other hand, if one starts directly with the steady-state equations by setting $\dfrac{\partial}{\partial t}=0$ in eqns. \ref{eqIVP-1}, one would have obtained the first term of the right hand side of eqn. \ref{eqIVP-7} as the particular integral of the problem. Since the coefficient of the left hand side of this has poles, the complete solution should also include delta function contributions corresponding to these poles. Accordingly, we write the solution obtained by setting $\dfrac{\partial}{\partial t}=0$ as:			
			\begin{eqnarray}\label{eqIVP-9}
				\dfrac{\eta(x)}{F_0} &=& - \dfrac{1}{2\pi}\int_{-\infty}^{\infty}dk \exp\left(ikx\right)\bigg\{\dfrac{1}{\alpha|k|^2 - \left(1+\rho_r\right)|k| + \left(1-\rho_r\right)}+ \nonumber \\
				&& + \; C_1 \delta(k-k_s) + C_2 \delta(k+k_s) + C_3 \delta(k-k_l) + C_4 \delta(k+k_l)\bigg\} \nonumber \\
				&=& -\;\dfrac{1}{\pi}\;\text{PV}\int_{0}^{\infty}dk\dfrac{\cos\left(kx\right)}{\alpha\left(k-k_l\right)\left(k-k_s\right)}
				+ c_1\cos(k_sx) + d_1\sin(k_sx)+ \nonumber \\
				&& +\; c_2\cos(k_lx) + d_2\sin(k_lx)
			\end{eqnarray}
		    where $c_1,d_1,c_2,d_2$ are (real) constants, $k_l$ and $k_s$ denote positive roots of the quadratic equation $\alpha |k|^2 - \left(1+\rho_r\right)|k| + \left(1-\rho_r\right)=0$ viz.
			\begin{equation}\label{eqIVP-10}
				k_{l,s} \equiv 	\dfrac{1+\rho_r}{2\alpha}\left(1\pm \sqrt{1-\dfrac{4\alpha\beta}{1+ \rho_r}}\right), \quad \mathbb{R}\left(k_l\right) > \mathbb{R}\left(k_s\right) > 0,
			\end{equation}		    
		    with $\mathbb{R}\left(\cdot\right)$ indicating real part of its argument and `PV' in \ref{eqIVP-9} denoting the Cauchy principal value of the integral. 
		    
		    A number of observations can be made concerning eqn. \ref{eqIVP-9}. First, the particular integral part of the solution in \ref{eqIVP-9} is symmetric in $x$, see fig. \ref{fig4a}.	    
		    This is contrary to observation in the same manner as the time-independent part of \ref{eqIVP-7} discussed earlier. However, the remaining four terms \textcolor{magenta}{involving delta functions} in eq. \ref{eqIVP-9} can introduce asymmetry, through appropriate choices of the unknown coefficients \textcolor{magenta}{$c_1,d_1,c_2,d_2$}. However, there is no apriori justification for such choices without invoking either observations or group-velocity arguments. Thus, the direct solution to the steady-state equations by itself cannot lead to the observed asymmetry. It is for these reasons that artifical viscosity within the steady-state framework, was introduced by \cite{rayleigh1883form}, generating the observed asymmetry as shown in fig. \ref{fig4b}. On the other hand, from the IVP perspective whether such an asymmetry can be introduced by the time-dependent term in eqn. \ref{eqIVP-7}, is a matter that needs exploration. Reverting back to the Cauchy principal-value integral \textcolor{magenta}{on the right hand side of} eqn. \ref{eqIVP-9}, it can be shown that for $0 < \alpha < \alpha_{\text{max}} \equiv  \dfrac{1}{4}\dfrac{\left(1+\rho_r\right)^2}{1- \rho_r}$ (super-critical regime):
			\begin{figure}
		    	\centering
		    	\subfloat[Integral \ref{eqIVP-11} \textcolor{magenta}{(Blue curve)}]{\includegraphics[scale=0.79]{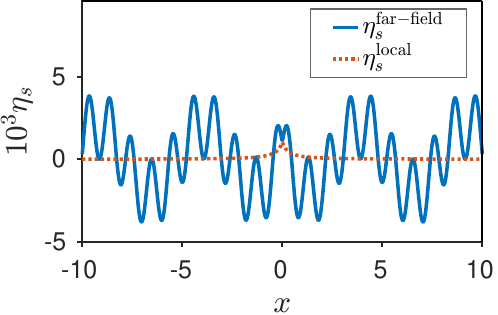}	\label{fig4a}\quad\quad}
		    	\subfloat[Expression \ref{eqIVP-12} \textcolor{magenta}{(Blue curve)}]{\includegraphics[scale=0.79]{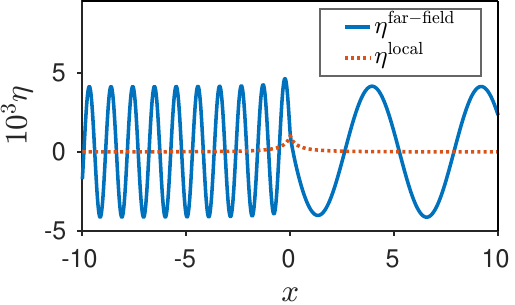}	\label{fig4b}} \\
    	     	\captionsetup{justification=raggedright,singlelinecheck=false}
		    	\caption{For both panels, $F_0=0.0014,\alpha=0.1389 < \alpha_{\text{max}}=0.25, \rho_r=0.001$. The roots $k_{l,s}$ are obtained from eqn. \ref{eqIVP-10}. Panel (a) The term $\eta_s^{\text{local}}(x)$ (see eqn. \ref{eqIVP-11}) has been evaluated in Matlab using the built-in function \texttt{integral}. Panel (b) Asymmetric upstream and downstream response (blue curve) as obtained from Rayleigh dissipation using expression \ref{eqIVP-12}. $\eta^{\text{far-field}}(x)$ and $\eta^{\text{local}}(x)$ are given by the terms with curly braces and the term involving $G(x)$ on the right hand side of eqn. \ref{eqIVP-12} respectively.}
		    	\label{fig4}
		    \end{figure}

		    \begin{eqnarray}\label{eqIVP-11}
		    	&& \dfrac{\eta_{s}(x)}{F_0}=-\dfrac{1}{\pi}\;\text{PV}\int_{0}^{\infty}dk\dfrac{\cos\left(kx\right)}{\alpha\left(k-k_l\right)\left(k-k_s\right)} =\dfrac{ \eta^{\text{far-field}}_{s}(x)}{F_0} + \dfrac{\eta^{\text{local}}_{s}(x)}{F_0}, \\		    	
		    	&&\text{with}\quad \dfrac{\eta^{\text{far-field}}_{s}(x)}{F_0} \equiv \dfrac{1}{\alpha(k_l-k_s)}\bigg\{-\sin(k_s|x|)\quad + \quad \sin(k_l|x|)\bigg\}, \nonumber \\
		    	&& \text{and}\quad\dfrac{\eta^{\text{local}}_{s}(x)}{F_0} \equiv  \dfrac{\left(k_l+k_s\right)}{\pi\alpha}\int_{0}^{\infty}dy \dfrac{y\exp\left(-|x|y\right)}{\left(y^2 + k_l^2\right)\left(y^2+k_s^2\right)},\quad x\neq 0\nonumber.
		    \end{eqnarray}
	        The partitioning of the response in eqn. \ref{eqIVP-11} into two terms viz. ``far-field'' and ``local'' (in space), serves to distinguish between the part which remains oscillatory sufficiently far from the forcing (at $x=0$) vis-a-vis the part which is significant only in a region around it. $\eta^{\text{local}}_s(x)\sim \mathcal{O}(1/|x|^2)$ \textcolor{magenta}{in eqn. \ref{eqIVP-11}}, for $|x| \rightarrow\infty$  which confirms its local nature. 
	        
        	The dissipation approach of \cite{rayleigh1883form}, used to generate fig. \ref{fig4b} is well-known \citep{lamb1932hydrodynamics}. In Appendix \ref{AppC}, we extend the approach to an interface between two fluids and demonstrate that this approach eliminates the need to determine these arbitrary constants $c_1,d_1,c_2,d_2$ in eqn. \ref{eqIVP-9}; the integrals involving these constants evaluate to zero. The positive dissipation term also displaces the poles $k_{s,l}$ in the integrand of eqn. \ref{eqIVP-9}, off the axis of integration in such a way that they contribute asymmetrically for $x >0$ and $x <0$. The dissipation is set to zero after applying the residue theorem. 
	        Interestingly, in Appendix \ref{AppC} we further prove that at any density ratio $0 \leq \rho_r < 1$, the integral \ref{eqIVP-9} when evaluated with positive \cite{rayleigh1883form} dissipation leads to the same functional form for $\eta(x)$ as derived by \cite{lamb1932hydrodynamics} but for $\rho_r=0$, see eqn. \ref{eqIVP-12} below. Thus, within the linearised approximation we find that the sole effect of change of $\rho_r$ beyond zero and for the super-critical regime with $0 < \alpha < \alpha_{\text{max}}$, is a quantitative shift in the (real) roots of the quadratic equation (above eqn. \ref{eqIVP-10}). The steady-state response $\eta(x)$ using the Rayleigh dissipation approach is obtained as:
			\begin{eqnarray}\label{eqIVP-12}
				&&\dfrac{\eta(x)}{F_0} = -\dfrac{2}{\alpha\left(k_{l} - k_{s}\right)}
				\left\{
				\begin{array}{lr}
					\sin(k_{l}x), & x<0\\
					\sin(k_{s}x), & x>0
				\end{array}
				\right\} + \dfrac{G(x)}{\pi\alpha}, \\
			 \text{with}\; &&G(x) \equiv \dfrac{1}{k_{l}-k_{s}}\int_{0}^{\infty}\;dk\;\left(\dfrac{\cos(kx)}{k+k_{s}}-\dfrac{\cos(kx)}{k+k_{l}}\right). \nonumber
			\end{eqnarray}
		    It is also shown in Supplementary Material (sec. $2.1$) that $\dfrac{G(x)}{\pi\alpha}$ in eqn. \ref{eqIVP-12} is identical to $\dfrac{\eta_s^{\text{local}}(x)}{F_0}$ in eqn. \ref{eqIVP-11}; the latter expression being preferable compared to $G(x)$ in \ref{eqIVP-12} due to the apparence of its local nature via the explicit exponential decay term \textcolor{magenta}{in the expression for $\eta^{\text{local}}_s(x)$ in eqn. \ref{eqIVP-11}}. Fig. \ref{fig4b} confirms that the steady-state response from the Rayleigh dissipation approach is qualitatively different from that of fig. \ref{fig4a}. Notably, the response employing Rayleigh dissipation is asymmetric about $x=0$ (fig. \ref{fig4b}), consistent with observations. In the next section, we show that similar results will be arrived at through the IVP approach.
		    
			\section{Analysis of the IVP solution}\label{sec:IVP-sol}
			We now turn to the time-dependent expressions in eqns. \ref{eqIVP-7} in the $\alpha = 0$ limit first. For inverting eqn. \ref{eqIVP-7} using \ref{eqIVP-8}, it is necessary to obtain the poles of the integrands. Note that $\alpha = 0$ is a singular limit in the sense that there is only one root to the otherwise quadratic eqn. $\alpha k^2 - \left(1+\rho_r\right)k + \left(1-\rho_r\right)=0$, see above eqn. \ref{eqIVP-10}. Recall that for $0 \leq \alpha < \alpha_{\text{max}}$ and $0 \leq \rho_r < 1$, there are always capillary and gravity waves up and downstream of the forcing, with wavenumbers $k_l$ and $k_s$ respectively (see expression \ref{eqIVP-10}). In the limit $\alpha\rightarrow 0$ and with $\rho_r <  1$, $k_l\rightarrow \infty$. Thus for $\alpha=0$ and $\rho_r <  1$, there is only one root viz. $k_s$ implying gravity waves downstream but no capillary ones upstream. \textcolor{magenta}{Consequently} with $\rho_r < 1$, the linear response for $\alpha \rightarrow 0$ and $\alpha=0$ are qualitatively different, in their upstream response ($x < 0$) but not in their downstream one. In \S\ref{subsec:alphaZero}, we consider $0 \leq \rho_r < 1$ and $\alpha=0$, interpreting this as the zero capillarity limit i.e. $T=0$ with gravity $g \neq 0$.
			\subsection{Zero capillarity limit: $\alpha = 0$}\label{subsec:alphaZero}
			For $\alpha=0$ (zero surface tension), expressions \ref{eqIVP-7} and \ref{eqIVP-8} may be written symbolically as (subscript `tr' indicates transient terms),
			\begin{eqnarray}\label{eqCan-1}
				\eta(x,t) &=& \eta_{s}(x) + \eta_{tr}^{(1)}(x,t) + \eta_{tr}^{(2)}(x,t), 
			\end{eqnarray}
			where
			\begin{subequations}\label{eqCan-2}
				\begin{align}	
					\dfrac{\eta_{s}(x)}{F_0} &\equiv \dfrac{1}{2\pi\left(1+\rho_r\right)}\int_{-\infty}^{\infty}dk\; \dfrac{\exp\left(ikx\right)}{|k| - \beta},\quad 0 < \beta \leq 1  \nonumber\\
					\dfrac{\eta_{tr}^{(1)}(x,t)}{F_0} &\equiv - \dfrac{1}{4\pi(1-\rho_r)}\int_{-\infty}^{\infty}dk\;\dfrac{k\;\exp\left[-i\left(k(t-x) + t\sqrt{\beta|k|}\right)\right]}{k+ \sqrt{\beta|k|}},  \nonumber\\
					\dfrac{\eta_{tr}^{(2)}(x,t)}{F_0} &\equiv - \dfrac{1}{4\pi(1-\rho_r)}\int_{-\infty}^{\infty}dk\;\dfrac{k\;\exp\left[-i\left(k(t-x) - t\sqrt{\beta|k|}\right)\right]}{k- \sqrt{\beta|k|}}. \tag{\theequation a,b,c}
					\end{align}
			\end{subequations}
			It may be further shown using principal value techniques that (see Supplementary Material, sec. $1.2$)
			\begin{eqnarray}\label{eqCan-3}
				\dfrac{\eta_{s}(x)}{F_0} &=& \dfrac{1}{\pi\left(1+\rho_r\right)}\left[-\pi\sin\left(\beta |x|\right)+\int_{0}^{\infty}dy\;\dfrac{y\exp\left(-|x|y\right)}{\beta^2 + y^2}\right],\quad 0 < \beta \leq 1,\; -\infty < x < \infty \nonumber \\
			\end{eqnarray}
			$\eta_{s}(x)$ in expression \ref{eqCan-3} is a symmetric function of $x$, implying a symmetric response upstream and downstream of the forcing at $x=0$. However, a contribution to the steady-state \textit{also comes} from the time-dependent term in in eqns. \ref{eqCan-2}(b),(c). As shown in the Supplementary Material (sec. $1.2$) these transient terms may be further simplified to obtain the following analytical representation valid for all $x,t$ i.e.	
			\begin{subequations}\label{eqCan-4}
				\begin{align}
				\eta_{tr}(x,t) &\equiv \eta_{tr}^{(1)}(x,t) + \eta_{tr}^{(2)}(x,t) = \bigg\{\mathbb{T}_1(x) + \mathbb{T}_2(x,t) + \mathbb{T}_3(x,t) + \mathbb{T}_4(x,t)\bigg\}F_0, \nonumber\quad\text{where} \\\nonumber\\
				\mathbb{T}_1(x)\equiv  \mp&\left(\dfrac{1}{1+\rho_r}\right)\sin(\beta x), \nonumber \\
				\mathbb{T}_2(x,t) \equiv &- \dfrac{4\beta^{-1}}{\pi\left(1+\rho_r\right)}\int_{0}^{\infty}dv\; v^2\;\dfrac{\exp\left(\mp 2av^2 \pm 2bv\right)}{\beta + \left(2v-\beta^{1/2}\right)^2}\Biggl\{\beta^{1/2}\cos\left(\beta^{1/2}tv\right)\pm \nonumber \\
				&\left(2v-\beta^{1/2}\right)\sin\left(\beta^{1/2}tv\right)\Biggr\}, \nonumber \\
				\mathbb{T}_3(x,t)\equiv &  \dfrac{\beta^{-1/2}}{\pi\left(1+\rho_r\right)}\left(1 + \dfrac{t}{2\left(t-x\right)}\right)\sqrt{\frac{\pi}{2|a|}}\left[\cos\left(\frac{b^2}{|a|}\right)\left\{\frac{1}{2}\mp \mathrm{C}\left(b\sqrt{\frac{2}{\pi |a|}}\right)\right\}+ \right. \nonumber\\
				&\left.\sin\left(\frac{b^2}{|a|}\right)\left\{\frac{1}{2}\mp \mathrm{S}\left(b\sqrt{\frac{2}{\pi |a|}}\right)\right\}\right], \nonumber \\
				\mathbb{T}_4(x,t)\equiv  &- \dfrac{1}{\pi\left(1+\rho_r\right)}\left[\int_{0}^{\infty}\;dv\dfrac{\cos\left(av^2 + 2bv\right)}{v + \beta^{1/2}}\right],
				\tag{\theequation a,b,c,d,e}	
			\end{align}
			\end{subequations}
     		where $a \equiv t-x, \; b \equiv \dfrac{t\sqrt{\beta}}{2}$, the upper signs in $\mathbb{T}_1(x),\mathbb{T}_2(x,t),\mathbb{T}_3(x,t)$ are used for $x<t$, while lower signs are for $x > t$. The Fresnel integrals $\mathrm{C}(\cdot)$ and $\mathrm{S}(\cdot)$ in eqn. \ref{eqCan-4}(d) are defined as
     		\begin{eqnarray}\label{eq:42s1gw}
     			\mathrm{C}\left(b\sqrt{\frac{2}{\pi |a|}}\right) \equiv  \bigintsss_{0}^{b\sqrt{\frac{2}{\pi |a|}}}dt\;  \cos\left(\frac{\pi t^2}{2}\right),\quad 
     			\mathrm{S}\left(b\sqrt{\frac{2}{\pi |a|}}\right) \equiv  \bigintsss_{0}^{b\sqrt{\frac{2}{\pi |a|}}}dt\;  \sin\left(\frac{\pi t^2}{2}\right). \nonumber
     		\end{eqnarray}
         
     	    In expressions \ref{eqCan-4}, the \textit{time-independent term},  $\mathbb{T}_1(x)$ (\ref{eqCan-4}(a)), is asymmetric with the same amplitude as the the first term on the right hand side of eq. \ref{eqCan-3}. As a result, these two terms reinforce each other for $x>0$ but cancel for $x<0$.  Further, we note that as $\rho_r\rightarrow 1$ $\left(\beta = \dfrac{1-\rho_r}{1 + \rho_r}\rightarrow 0\right)$, the terms diverge. This is physically reasonable because in this limit ($\rho_r\rightarrow 1$), gravity vanishes and in the absence of capillary forces as well (i.e. $\alpha=0$ that we are currently assuming), there remains no restoring force to resist deformation due to the external pressure. The analytical strategy is clear now: provided one can show that $\mathbb{T}_2(x,t\rightarrow\infty)\rightarrow0,\mathbb{T}_3(x,t\rightarrow\infty)\rightarrow0$ and $\mathbb{T}_4(x,t\rightarrow\infty)\rightarrow0$, one obtains the expected steady-state lacking waves upstream (except for small localised deformation of the interface due to the localised integral in eqn. \ref{eqCan-3}) and sinusoidal waves downstream ($x>0$) with wavenumber $\beta$.
     	    \subsubsection{Asymptotic form of the transient integral(s) for $t\rightarrow\infty$:}\label{subsubsec:asymp}
     	    \textcolor{magenta}{In order to see the transient behaviour of the time-dependent integrals $\mathbb{T}_2(x,t),\mathbb{T}_3(x,t)$ and $\mathbb{T}_4(x,t)$, it is useful to revert to eqns. 
     	    	\ref{eqCan-2}(b),(c). Refer to online supplementary material (S.M.) \citep{Kadari2026supplementary}, these two integrals in eqns. \ref{eqCan-2}(b),(c) (without the pre-factors) may be re-written as $\int_0^\infty
     	    	\frac{k}{k+\sqrt{\beta k}}
     	    	\cos\left\{
     	    	t\left(k+\sqrt{\beta k}\right)-kx
     	    	\right\}\,dk$ and $\int_0^\infty
     	    	\frac{k}{k-\sqrt{\beta k}}
     	    	\cos\left\{
     	    	t\left(k-\sqrt{\beta k}\right)-kx
     	    	\right\}\,dk$ respectively. It may then be shown using the method of stationary-phase (see online S.M. \citep{Kadari2026supplementary}, section `Theory$\rightarrow$ Pure Gravity$\rightarrow$Asymptotic form of the$\ldots$') that for $t\rightarrow\infty$:}
     	    \begin{subequations}\label{eqAsymp}
     	    	\begin{align}
     	    	&\int_0^\infty
     	    	\frac{k}{k+\sqrt{\beta k}}
     	    	\cos\left\{
     	    	t\left(k+\sqrt{\beta k}\right)-kx
     	    	\right\}\,dk \sim -\left(\dfrac{60}{\beta^3}\right)t^{-4} \nonumber \\
     	    	& \int_0^\infty
     	    	\frac{k}{k-\sqrt{\beta k}}
     	    	\cos\left\{
     	    	t\left(k-\sqrt{\beta k}\right)-kx
     	    	\right\}\,dk  \sim     	    	
     	    	2\pi\beta\sin(\beta x)
     	    	-\sqrt{\pi\beta}\;t^{-1/2}\cos\left[
     	    	\frac{\beta(t+x)}{4}
     	    	-\frac{\pi}{4}
     	    	\right]  \nonumber \\
     	    	\tag{\theequation a,b}
     	    	\end{align}  	
     	    \end{subequations}
         \textcolor{magenta}{It is clear from these expressions that the slower decay is of the form $t^{-1/2}$. In the online S. M. \citep{Kadari2026supplementary}, comparisons are these asymptotic estimates alongwith numerical solutions to the integrals on the left hand side of eqns. \ref{eqAsymp}(a),(b) are also provided.}
         
\begin{figure}
            	\centering
            	\subfloat[$t=0.37$]{\includegraphics[width=0.47\textwidth]{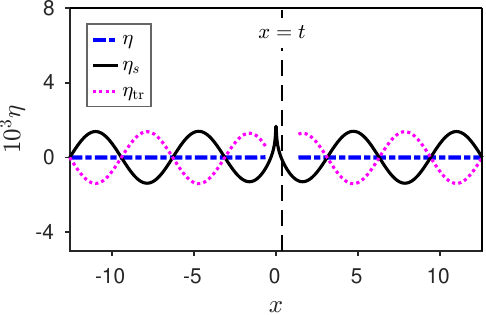} \label{fig5a}}
            	\hspace{0.04\textwidth}
            	\subfloat[$t=2.57$]{\includegraphics[width=0.47\textwidth]{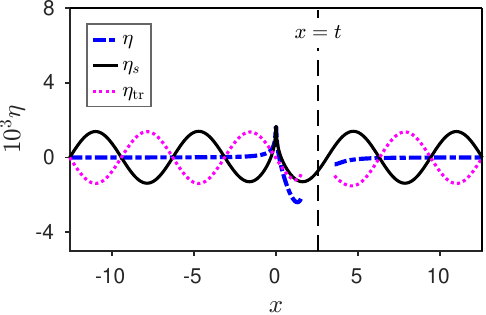} \label{fig5b}} \\
            	\subfloat[$t=5.51$]{\includegraphics[width=0.47\textwidth]{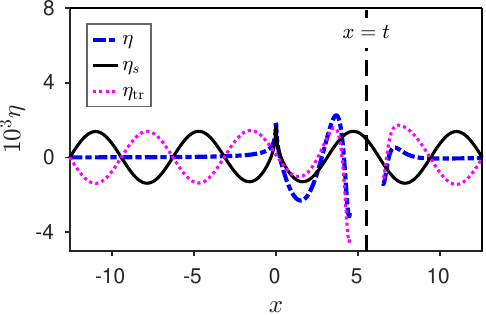} \label{fig5c}}
            	\hspace{0.04\textwidth}
            	\subfloat[$t=14.69$]{\includegraphics[width=0.47\textwidth]{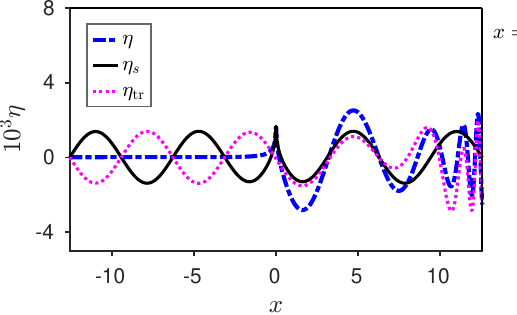} \label{fig5d}}\\
            	\subfloat[$t=22.04$]{\includegraphics[width=0.47\textwidth]{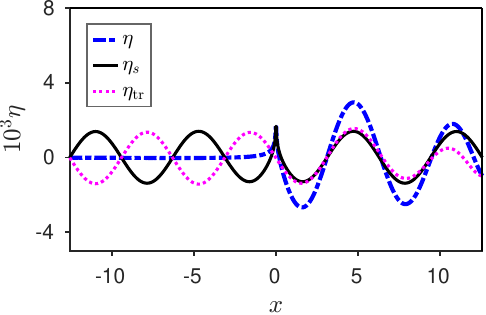} \label{fig5e}}
            	\hspace{0.04\textwidth}
            	\subfloat[$t=36.74$]{\includegraphics[width=0.47\textwidth]{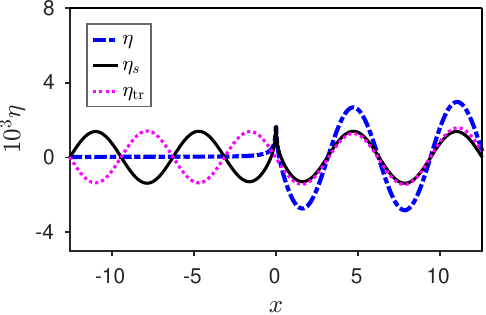} \label{fig5f}}\\      
            	\subfloat[$t=110.21$]{\includegraphics[width=0.47\textwidth]{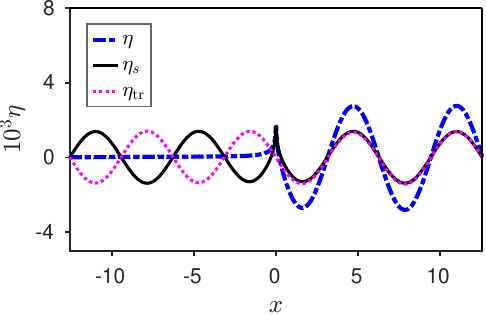} \label{fig5g}}
            	\hspace{0.04\textwidth}
            	\subfloat[$t=183.68$]{\includegraphics[width=0.47\textwidth]{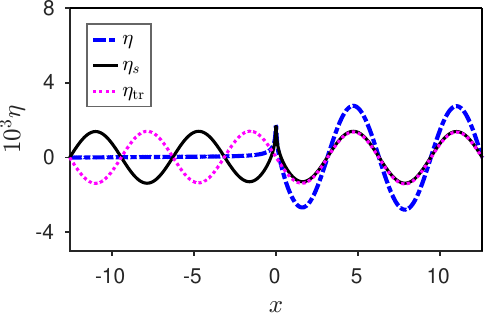} \label{fig5h}}
       	     	\captionsetup{justification=raggedright,singlelinecheck=false}
            	\caption{Time evolution of $\eta$ $\left(\rho_r=0.001, \alpha=0, F_{0}=0.0014\right)$. Here $\eta_{s}(x)$ (see eqn. \ref{eqCan-3}) is the time-independent part of $\eta(x,t)$ (eqn. \ref{eqCan-1}), $\eta_{tr}(x,t)$ (eqns. \ref{eqCan-4}(a)-(e)) is the transient part. Vertical dashed line in panels (a,b,c) are at $x=t$ with most of the transient response seen for $x < t$, see panel (c). The integrals (eqns. \ref{eqCan-3} and \ref{eqCan-4}) are evaluated using \citep{MATLAB:2022b} in-built function \texttt{integral}. The upper limit when infinite, are replaced by $10^2$ ensuring that results do not change on this scale with further increase.}
            	\label{fig5}
            \end{figure}

         Fig. \ref{fig5} presents the time evolution of the interface ($\eta$ in eqn. \ref{eqCan-1}), the steady part $\eta_{s}(x)$ (eqn. \ref{eqCan-2}(a)) and the transient part (eqn. \ref{eqCan-4}(a)). In the upstream region $x < 0$, a clear cancellation between $\eta_s(x)$ and $\eta_{tr}(x,t)$ is seen. In the absence of this cancellation, the steady response $\left(\eta_s(x)\right)$ would be completely symmetric about $x=0$. Note that the term $\mathbb{T}_2(x,t)$ diverges as $x\rightarrow t$. In dimensional variables, this is the location of the fluid parcel which was at $\tilde{x}=0$ initially $\left(\tilde{t}=0\right)$. The special nature of this point $x=t$, was also noted in the work of \cite{wurtele1955transient} (also for $\alpha=0$) where for $x=t$ (in scaled coordinates), the saddle-point of the integrand $\rightarrow\infty$, see expression $3.7$ in \cite{wurtele1955transient}. To prevent this divergence in our plots, we have excluded a small region around $x=t$ in fig. \ref{fig5}.
          
			\subsection{Finite capillarity: $\alpha > 0$}\label{subsec:AlphGT0}
			We now turn to the case of $\alpha> 0$. As $\rho_r < 1$, we have both capillary and gravitational forces now. For this case, we may write similar to the previous sub-section (see Supplementary Material, sec. $1.1$) i.e.
\begin{subequations}\label{eqCan-5}
	\begin{align}
		&\eta(x,t) = \eta_{s}(x) + \eta_{tr}(x,t), \nonumber \\ \nonumber \\
		&\dfrac{\eta_s(x)}{F_0} \equiv  -\dfrac{1}{\pi}\int_{0}^{\infty} dk\;
		\left\{\frac{\cos(kx)} 
		{\alpha\,(k-k_l)(k-k_s)}\right\},\quad 
		\dfrac{\eta_{\text{tr}}(x,t)}{F_0} 
		\equiv -\dfrac{1}{2\pi}\bigg\{\mathbb{I}_3(x,t) + \mathbb{I}_4(x,t)\bigg\}, \nonumber \\\nonumber \\
		&\mathbb{I}_{3,4}(x,t) \equiv   -\dfrac{(1+\rho_r)}{\alpha}\int_{0}^{\infty}dk\;\dfrac{\left(k\pm\chi(k)\right)\cos\left[t\left(k\mp\chi(k)\right)-kx\right]}{\left(1 + \alpha k^2 - \rho_r\right)\left(k-k_l\right)\left(k-k_s\right)},\;\;\chi(k) \equiv \sqrt{\beta k+\frac{\alpha}{1+\rho_r}\,k^3}, 
		\tag{\theequation a,b,c,d}
	\end{align}
\end{subequations}
where $k_l,k_s$ are from eqn. \ref{eqIVP-10}. Similar to the previous section, the interface shape due to the time-independent response $\eta_{s}(x)$ (eqn. \ref{eqCan-5}(b)), here too, is symmetric about $x=0$ as seen in the first panel (a) of fig. \ref{fig7}, see black solid curve.

\begin{figure}
	\centering
   	\captionsetup{justification=raggedright,singlelinecheck=false}
	\includegraphics[scale=0.8]{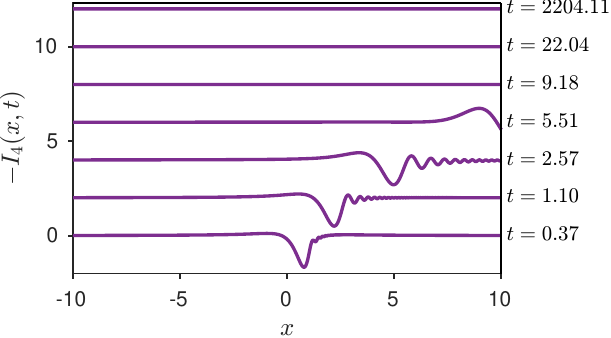}
	\caption{Time evolution of $\mathbb{I}_4(x,t)$ from eqn. \ref{eqCan-5}(d) for $\rho_r=0.001$ and $\alpha=0.1389$. Each curve (except the first curve at $t=0.37$) is shifted vertically upwards by two units compared to the previous one.}
	\label{fig6}
\end{figure}

We now turn to a formal demonstration of the asymmetric cancellations (about $x=0$) as $t\rightarrow\infty$. The expression for $\eta_{s}(x)$ in eqn. \ref{eqCan-5}(b), after application of principal-value techniques may be written as (see Supplementary Material sec. $1.3$ \textcolor{magenta}{or online S.M. \citep{Kadari2026supplementary} for the proof}):
\begin{equation}\label{eqCan-6}
	\dfrac{\eta_{s}(x)}{F_0} = \dfrac{1}{\alpha(k_l-k_s)}\bigg\{-\sin(k_s|x|) + \sin(k_l|x|)\bigg\} + \left(\dfrac{k_l+k_s}{\alpha\pi}\right)\int_{0}^{\infty}dy \dfrac{y\exp\left(-|x|y\right)}{\left(y^2 + k_l^2\right)\left(y^2+k_s^2\right)}.
\end{equation}
   
   After lengthy calculations involving contour integration and stationary-phase approximation (presented in the Supplementary Material, sec. $1.3$), we may show that 
   \begin{eqnarray}\label{eqCan-7}
   		\dfrac{\eta_{\text{tr}}(x,t\rightarrow\infty)}{F_0}
   		=\frac{1}{\alpha(k_l-k_s)}\bigg(-\sin(k_s x)-\sin(k_l x)\bigg),\quad x\in \left(-\infty,\infty\right).
   \end{eqnarray}
This contribution  to the steady-state essentially stems from the term $\mathbb{I}_3(x,t)$ in eqn. \ref{eqCan-5}(d) whereas the term $\mathbb{I}_4(x,t)$ in the same equation tends to zero as $t\rightarrow\infty$. Fig. \ref{fig6} confirms this decay for large time ($t >> 1$). The sum of \ref{eqCan-6} and \ref{eqCan-7} yields the final form of the steady-state interface at all $x$.
The asymmetric cancellation in $\eta(x,t\rightarrow\infty) = \eta_s(x) + \eta_{\text{tr}}(x,t\rightarrow\infty)$, upstream ($x < 0$) and downstream ($x > 0$) of the forcing, may readily be observed by comparing these expressions. We reiterate that the short (for $x < 0$) and long (for $x > 0$) waves seen at steady-state, result from this cancellation. Unlike the $\alpha=0$ case, it was not possible to obtain closed form expressions in terms of real integrals for the $\eta_{tr}(x,t)$ terms in eqns. \ref{eqCan-5}(d).  Hence, we have evaluated these integrals directly numerically in the principal-value sense around the pole(s). Results for the time evolution based on these, are presented in Fig. \ref{fig7} for $\alpha = 0.1389,\rho_r=10^{-3}$. The asymmetric cancellations leading to the steady-state in panel (h) of the figure are apparent.
Notice in particular, panels (g) and (h) in fig. \ref{fig7} where $\eta_{\text{tr}}(x,t)$ does not change between these two panels; this is the time-independent contribution from $\eta_{\text{tr}}(x,t)$. 

\begin{figure}
	\centering
	\subfloat[$t=0.37$]{\includegraphics[width=0.47\textwidth]{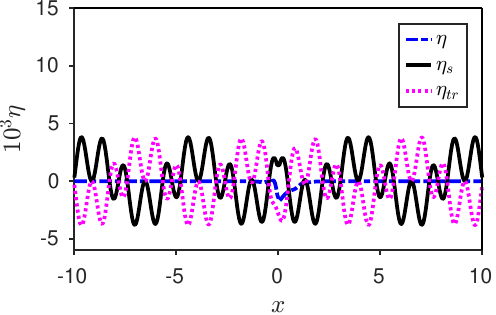} \label{fig7a}}
	\hspace{0.04\textwidth}
	\subfloat[$t=2.57$]{\includegraphics[width=0.47\textwidth]{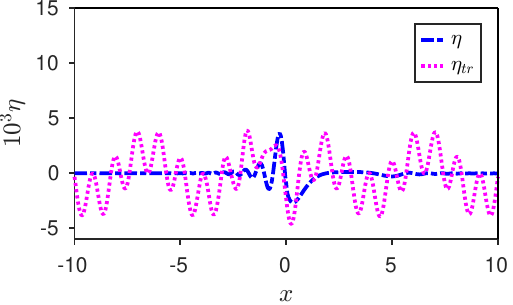} \label{fig7b}} \\
	\subfloat[$t=5.51$]{\includegraphics[width=0.47\textwidth]{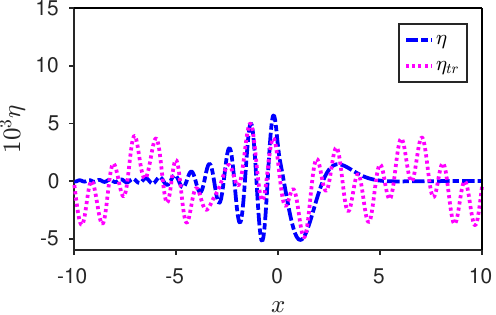} \label{fig7c}}
	\hspace{0.04\textwidth}
	\subfloat[$t=14.69$]{\includegraphics[width=0.47\textwidth]{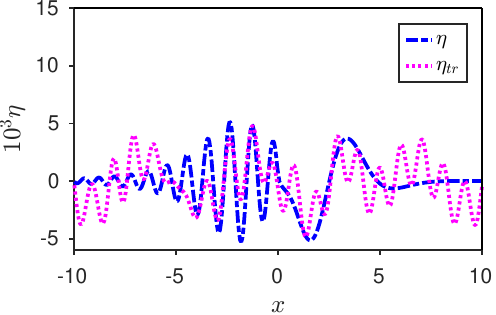} \label{fig7d}}\\
	\subfloat[$t=22.04$]{\includegraphics[width=0.47\textwidth]{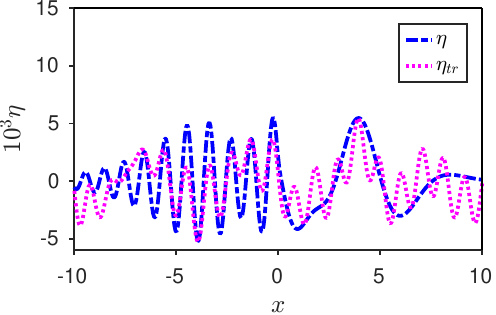} \label{fig7e}}
	\hspace{0.04\textwidth}
	\subfloat[$t=36.74$]{\includegraphics[width=0.47\textwidth]{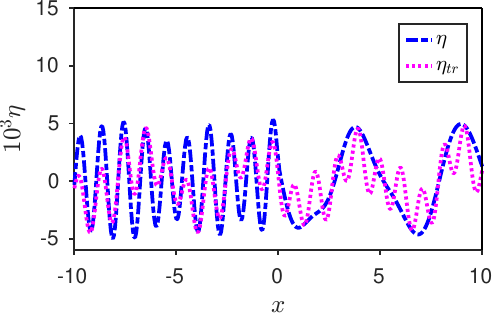} \label{fig7f}}\\      
	\subfloat[$t=367.35$]{\includegraphics[width=0.47\textwidth]{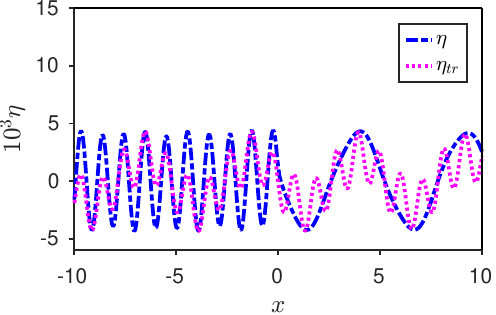} \label{fig7g}}
	\hspace{0.04\textwidth}
	\subfloat[$t=2204.1$]{\includegraphics[width=0.47\textwidth]{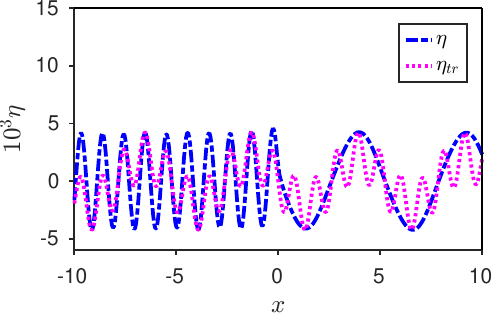} \label{fig7h}}
 	\captionsetup{justification=raggedright,singlelinecheck=false}
	\caption{Time evolution for $\rho_r=10^{-3}, \alpha=0.1389, F_{0}=0.0014$ as obtained by numerically solving expressions \ref{eqCan-5}.}
	\label{fig7}
\end{figure}

We next turn to establishing the validity of above linearised, time-dependent theory vis-a-vis nonlinear simulations of the incompressible, Euler's equations. Such benchmarking tests with analytical time-dependent theory have not been reported so far. \textcolor{magenta}{An additional novelty that will be apparent from the nonlinear simulations is that the localised pressure forcing applied on an uniformly moving layer of liquid, appears as a plausible mechanism to generate moderately steep, Stokes or Crapper \citep{Crapper1957} waves in the laboratory.}

	\section{Comparison with nonlinear simulations}\label{sec:nonlin_sim}
\begin{figure}
		\centering
		\subfloat[Domain]{\includegraphics[scale=0.9]{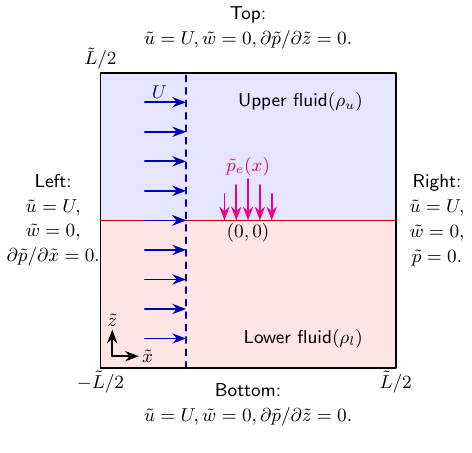}\label{fig8a}}\quad
		\subfloat[Adaptive grid]{\includegraphics[scale=0.24]{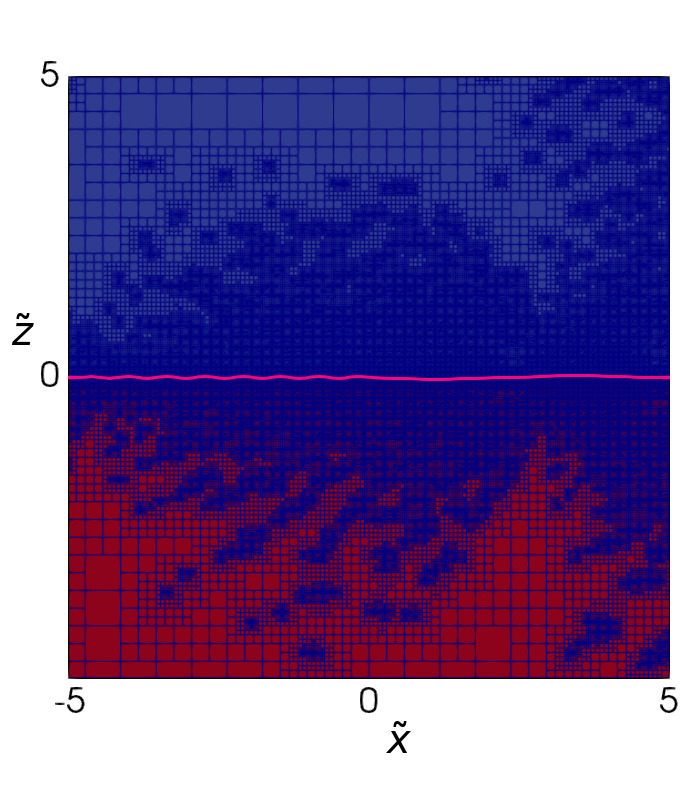}\label{fig8b}}
	   	\captionsetup{justification=raggedright,singlelinecheck=false}
		\caption{\textcolor{magenta}{Panel (a) A (not to scale) schematic of the simulation domain. The interface is initially flat and at $\tilde{z}=0$. A local pressure forcing $\tilde{p}_e(x, t>0)= \dfrac{\tilde{F}_0 }{\pi}\dfrac{\tilde{b}}{(\tilde{b}^2 + \tilde{x}^2)}$ with $\tilde{b}=9.2\times10^{-5}$ m (overpressure $\tilde{p}_e(x,t)\rightarrow\tilde{F_0}\delta(\tilde{x})$ as $\tilde{b}\rightarrow 0$) is applied at $\tilde{x}=0,\tilde{z}=0$ for $\tilde{t}>0$. Initially ($\tilde{t}=0$), an uniform horizontal velocity $U\approx 0.267\; \text{m/s}$ and zero vertical component, is imposed on all cells in the computational domain (both fluids). With $\rho_l=10^3$ kg/m$^3$, $T=0.072$ N/m and $g=9.81$ m/s$^2$, this choice of $U$ corresponds to $\alpha \equiv \dfrac{gT}{\rho_lU^4}=0.1389 < \alpha_{\text{max}}$. In all simulations $\rho_l=10^3\; \text{kg}/\text{m}^3$ while $\rho_r\in \left[10^{-3},10^{-1}\right]$. Panel (b) Sample adaptive grid with interface (pink).}}
		\label{fig8}
	\end{figure}

	In this section, we compare our linearised time-dependent predictions with nonlinear simulations. The latter are conducted using the open-source code Basilisk \citep{popinet2025basilisk}. \textcolor{magenta}{As the benchmark in appendix \ref{AppF-Basilisk} demonstrates via the temporal evolution of a nonlinear Stokes wave, Basilisk is an accurate computational framework for potential flow problems with a free-surface.} A schematic of the simulation domain with boundary and initial conditions are presented in fig. \ref{fig8a} while fig. \ref{fig8b} shows a sample adaptive grid used. \textcolor{magenta}{The domain length $\tilde{L}\approx0.756\;$ m is held constant in all simulations, several long and short waves may fit within this domain in the chosen parameter regime. The adaptive mesh refinement (fig. \ref{fig8b}) employs maximum and minimum grid resolution of $2^{13}$ and $2^{5}$ respectively, implying smallest and largest spatial resolutions, $\tilde{L}/2^{13}$ and $\tilde{L}/2^{5}$ respectively. This is based on change in volume fraction $f$, velocity $\mathbf{u}$ and interface curvature $\kappa$. Curvature based adaptivity is employed only in a reduced part of the simulational domain $\left[-0.3,\;0.3\right]\;\text{m}$, in order to create a \textit{buffer region} extending up to the computational boundary at $\tilde{x}=\pm \tilde{L}/2$ and restricting reflections from the same (see further details in section \ref{subsubsec:SimApprox}).}

	 \textcolor{magenta}{We note that} the open-source code Basilisk has been extensively used by the two-phase flow research community over the last decade, benchmarking it against a large number of analytical solutions as well as experimental observations, particularly in the large Reynolds number regime; these include unsteady interfacial flow problems \citep{sanjay2023does}, capillary-gravity waves \citep{yang2026surfactant}, contact lines \citep{fullana2026consistent} etc. \textcolor{magenta}{Also note that Basilisk \citep{popinet2025basilisk} has been extensively used to investigate strongly nonlinear waves e.g. for studying wave-breaking with good match being reported against experimental results see e.g. fig. $13$ in \cite{Deike2016Airentrainment} or more recently in \cite{Lande2026Validation} (their section $3$ and fig. $6$(b))}.	 
	 Within our group, we have used Basilisk (or Gerris) extensively to solve the incompressible, Euler's or Navier-Stokes equation with an interface accounting for gravity and/or surface-tension, obtaining excellent agreement with theoretical predictions in linear and nonlinear, viscous and inviscid regimes in a variety of interfacial flow problems \citep{farsoiya2017axisymmetric,patankar2018faraday, singh2019test, kayal2022dimples, patankar2022dynamic, kayal2023jet, patibandla2023surface, dhote2024standing, kayal2025focussing,kumar2025waves}. 
	 
	 The code numerically solves the incompressible, fluid flow equations including gravity and surface-tension employing the volume-of-fluid algorithm \citep{tryggvason2011direct} to represent the interface of our interest here. The equations which are numerically solved are the mass and momentum equation (Euler's equation) and \textcolor{magenta}{an} advection equation for transport of a volume-fraction field. These equations are well-known and reproduced here for completeness:
	\begin{subequations}\label{eqSim-1}
		\begin{align}
			& \bm{\tilde{\nabla}.\tilde{u}}=0,\quad \dfrac{\partial\bm{\tilde{u}}}{\partial \tilde{t}}+\bm{\tilde{\nabla}}.(\mathbf{\tilde{u}\otimes \tilde{u}}) = - \dfrac{\bm{\tilde{\nabla}}\tilde{p}}{\rho} + \mathbf{\tilde{g}} + \dfrac{T}{\rho}\kappa \delta_s \bm{n} + \dfrac{\tilde{p}_e (\tilde{x})}{\rho} \left(\dfrac{\partial f}{\partial \tilde{z}}\right) \hat{j} ,\label{eqSim-1b} \\
			& \text{and}\quad\dfrac{\partial f}{\partial \tilde{t}}+\bm{\tilde{\nabla}.}(f\bm{\tilde{u}})=0. \label{eqSim-1c}
		\end{align}
	\end{subequations}
	Here $\mathbf{\tilde{u}},\,\tilde{p},\,\mathbf{\tilde{g}},\,\kappa$, $T$ and $f$ are the velocity field, pressure field, gravity, interface curvature, surface-tension and the volume-fraction field respectively while $\delta_s$ is a surface delta function and $\hat{j}$ is an unit vector in the vertically upward direction. 
	
	The last term on the right hand side of eqn. \ref{eqSim-1}(a) models the overpressure. This term is restricted to the interface by multiplying $\tilde{p}_e(\tilde{x})$ with the volume-fraction gradient $\dfrac{\partial f}{\partial \tilde{z}}$. Since $f$ changes sharply only across the interface, $\dfrac{\partial f}{\partial \tilde{z}}$ is non-zero only in the interfacial region. Consequently, the forcing acts only at the interface.  This imposed acceleration may be interpreted as $
	a_z \approx \dfrac{1}{\rho}\,\tilde{p}_e(\tilde{x})\,\dfrac{\partial f}{\partial \tilde{z}}$. 	In the present simulations, the volume-fraction field is such that the this (pressure) force acts downwards on the interface. The function $\tilde{p}_e(\tilde{x})$ is chosen of the Lorentzian form with width  $\tilde{b}=\frac{\tilde{L}}{2^{13}}=9.2\times10^{-5}$ m. This is \textcolor{magenta}{approximately one computational cell} wide near the interface, so that $\tilde{b}$ is small and thus the pressure form approaches a Dirac delta-function, permitting comparison with \textcolor{magenta}{linear} theory. \textcolor{magenta}{The Basilisk script used for simulations is provided in \cite{Kadari2026supplementary}, see section titled `Basilisk setup for forced Capillary-Gravity waves' for details.}
	\subsection{\textcolor{magenta}{The buffer region in nonlinear simulations}}\label{subsubsec:SimApprox}	
	While our \textcolor{magenta}{linearised} theory is for a horizontally unbounded configuration with $x \in \left(-\infty,\infty\right)$, the simulation domain is bounded between $\tilde{x}\in \left[-\tilde{L}/2,\tilde{L}/2\right]$ with $\tilde{L}= 0.756$ m. \textcolor{magenta}{There is a significant body of literature on numerical implementation of radiation boundary conditions in transient wave problems, using techniques such as perfectly matched layers (PML) \citep{Berenger1994PML}, absorbing boundary conditions \citep{Engquist1977Absorbing} etc\footnote{We thank an anonymous referee for pointing these techniques to us}. We have however avoided imposition of these in our nonlinear simulations, to remain consistent with our assertion that the IVP approach does not require radiation conditions.} As a consequence of the boundary conditions applied at $\tilde{x}=\pm \tilde{L}/2$ (see fig. \ref{fig8a} for details) in simulations \textcolor{magenta}{however}, wave reflections \textcolor{magenta}{occur} when the simulation is continued for sufficiently long time. \textcolor{magenta}{To circumvent these reflections not described in our theory, we provide a buffer region ( $0.078$ m in length) for both left and right simulation boundaries; in this buffer zone a grid density coarser than the domain of our interest is employed to weaken these reflections.} Comparisons with linear theory and nonlinear simulations are discussed in the following sub-sections.
	\subsection{Linear regime} \label{subsec:lin}
\begin{figure}
     	\centering
     	\subfloat[$t=0.37$]{\includegraphics[width=0.47\textwidth]{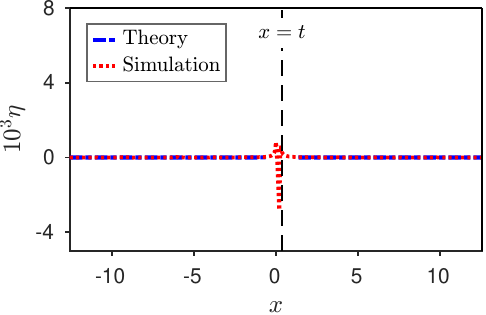} \label{fig9ga}}
     	\hspace{0.04\textwidth}
     	\subfloat[$t=2.57$]{\includegraphics[width=0.47\textwidth]{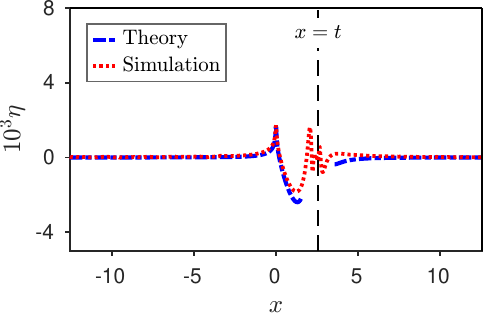} \label{fig9gb}} \\
     	\subfloat[$t=5.51$]{\includegraphics[width=0.47\textwidth]{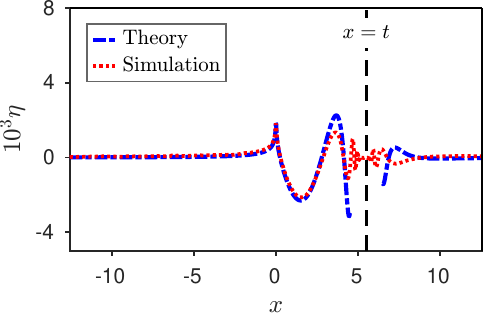} \label{fig9gc}}
     	\hspace{0.04\textwidth}
     	\subfloat[$t=14.69$]{\includegraphics[width=0.47\textwidth]{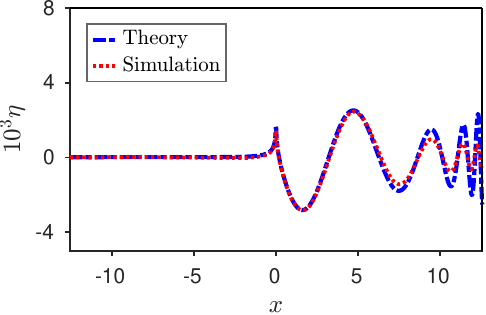} \label{fig9gd}}\\
     	\subfloat[$t=22.04$]{\includegraphics[width=0.47\textwidth]{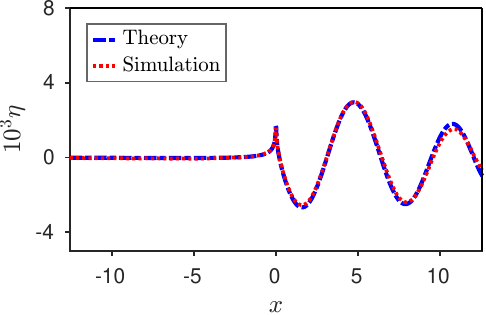} \label{fig9ge}}
     	\hspace{0.04\textwidth}
     	\subfloat[$t=36.74$]{\includegraphics[width=0.47\textwidth]{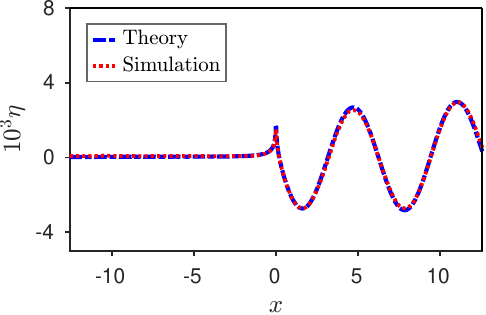} \label{fig9gf}}\\      
     	\subfloat[$t=110.21$]{\includegraphics[width=0.47\textwidth]{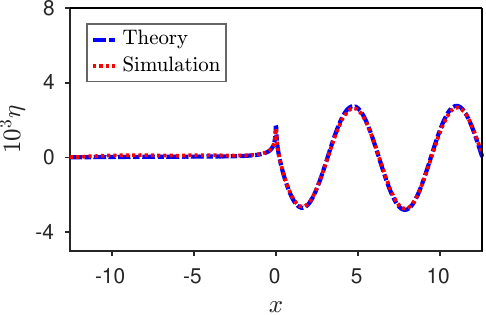} \label{fig9gg}}
     	\hspace{0.04\textwidth}
     	\subfloat[$t=183.68$]{\includegraphics[width=0.47\textwidth]{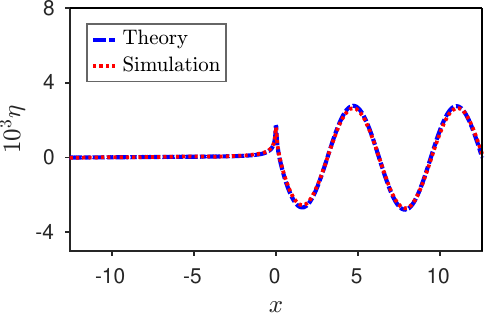} \label{fig9gh}}
     	\captionsetup{justification=raggedright,singlelinecheck=false}
     	\caption{\textcolor{magenta}{Comparison between linear theory (dashed-dot blue curve) (IVP in section \ref{subsec:alphaZero}) for $\alpha=0,\rho_r=10^{-3}, F_0=1.4\times10^{-3}$ and nonlinear simulations (red dotted curve) at various instants of time.}}
     	\label{fig9}
     \end{figure}

\begin{figure}
     	\centering
     	\subfloat[$t=0.37$]{\includegraphics[width=0.47\textwidth]{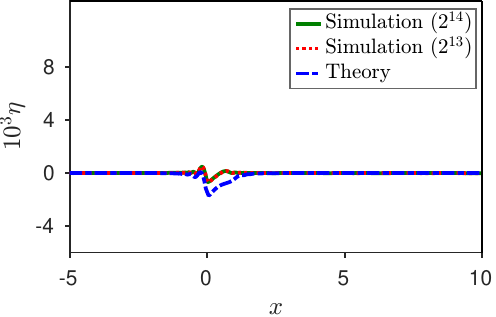} \label{fig10a}}
     	\hspace{0.04\textwidth}
     	\subfloat[$t=1.10$]{\includegraphics[width=0.47\textwidth]{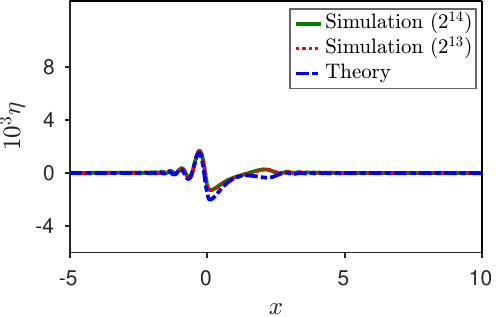} \label{fig10b}} \\
     	\subfloat[$t=2.57$]{\includegraphics[width=0.47\textwidth]{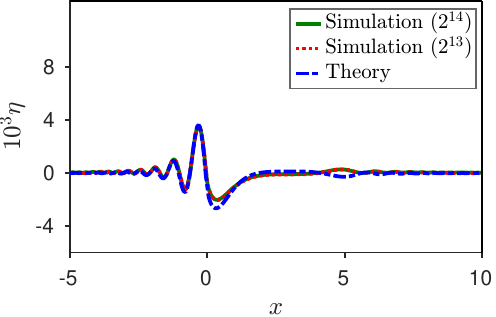} \label{fig10c}}
     	\hspace{0.04\textwidth}
     	\subfloat[$t=5.51$]{\includegraphics[width=0.47\textwidth]{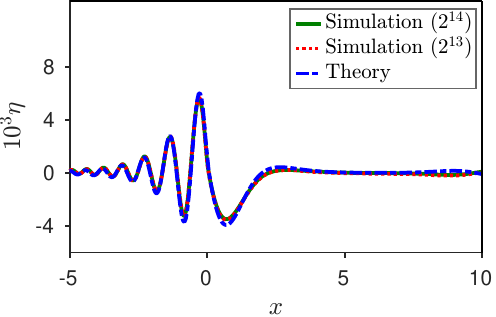} \label{fig10d}}\\
     	\subfloat[$t=9.18$]{\includegraphics[width=0.47\textwidth]{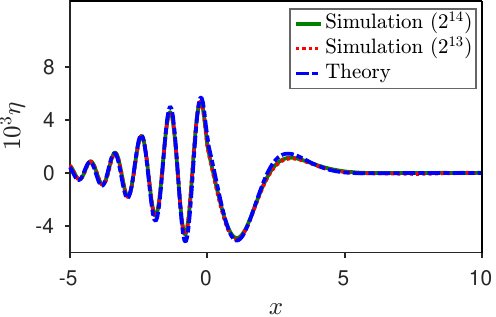} \label{fig10e}}
     	\hspace{0.04\textwidth}
     	\subfloat[$t=22.04$]{\includegraphics[width=0.47\textwidth]{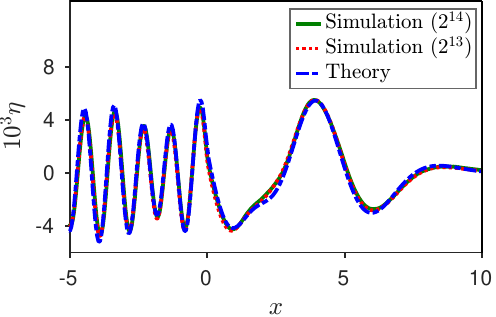} \label{fig10f}}\\      
     	\subfloat[$t=53.27$]{\includegraphics[width=0.47\textwidth]{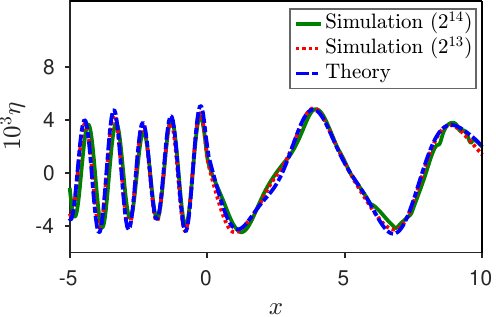} \label{fig10g}}
     	\hspace{0.04\textwidth}
     	\subfloat[$t=110.21$]{\includegraphics[width=0.47\textwidth]{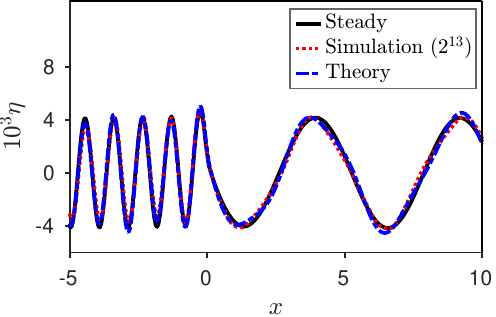} \label{fig10h}}
     	\captionsetup{justification=raggedright,singlelinecheck=false}
     	\caption{Comparison of linear theory (eqns. \ref{eqCan-5}(a)-(d)) with simulations for $\rho_r=10^{-3},\alpha=0.1389$ and $F_0\equiv \dfrac{\tilde{F}_0}{\rho_l U^2 l_c}=1.4\times 10^{-3}$ with length and time in units of $l_c\equiv U^2/g$ and $t_c \equiv U/g$ respectively. The solid black curve in panel (h) labelled as `Steady' is from expression \ref{eqIVP-12}.}
     	\label{fig10}
     \end{figure}

\begin{figure}
     	\centering
     	\subfloat[$t=0.37$]{\includegraphics[width=0.47\textwidth]{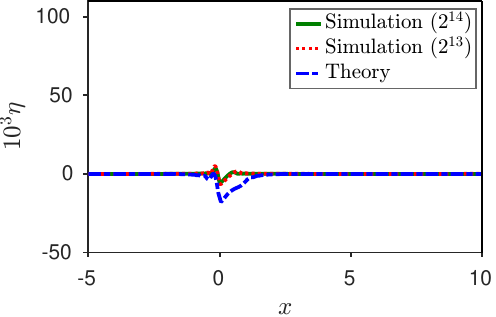} \label{fig11a}}
     	\hspace{0.04\textwidth}
     	\subfloat[$t=1.10$]{\includegraphics[width=0.47\textwidth]{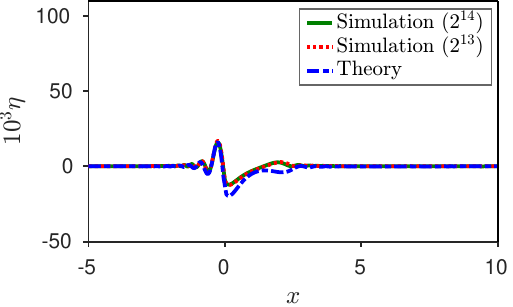} \label{fig11b}} \\
     	\subfloat[$t=2.57$]{\includegraphics[width=0.47\textwidth]{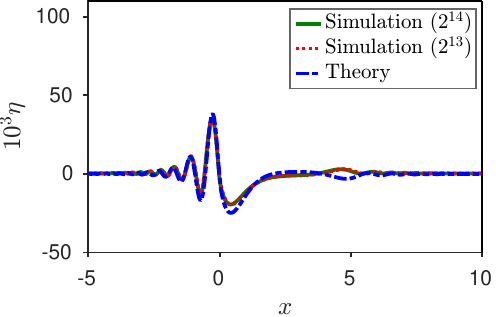} \label{fig11c}}
     	\hspace{0.04\textwidth}
     	\subfloat[$t=5.51$]{\includegraphics[width=0.47\textwidth]{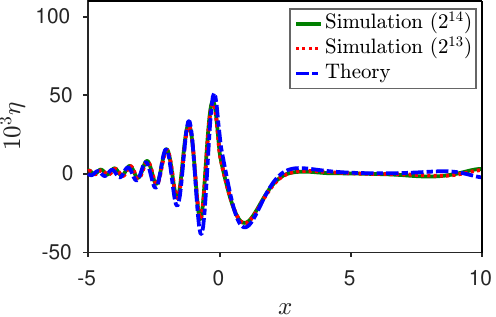} \label{fig11d}}\\
     	\subfloat[$t=14.69$]{\includegraphics[width=0.47\textwidth]{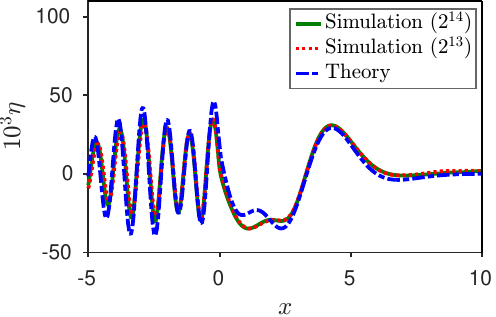} \label{fig11e}}      
     	\hspace{0.04\textwidth}
     	\subfloat[$t=18.37$]{\includegraphics[width=0.47\textwidth]{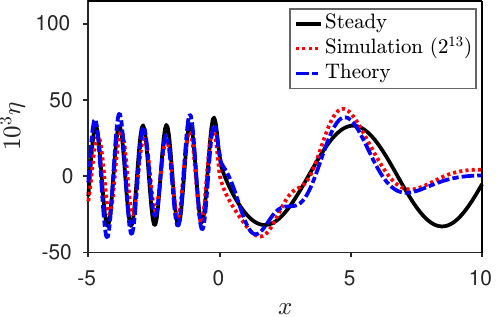} \label{fig11f}}
     	\captionsetup{justification=raggedright,singlelinecheck=false}
     	\caption{Comparison of linear theory (eqns. \ref{eqCan-5}(a)-(d)) with simulations for $\rho_r=10^{-1},\alpha=0.1389$ and $F_0\equiv \dfrac{\tilde{F}_0}{\rho_l U^2 l_c}=1.4\times 10^{-2}$. The solid black curve in panel (f) labelled as `Steady' is from expression \ref{eqIVP-12}.}
     	\label{fig11}
     \end{figure}

	\textcolor{magenta}{Fig. \ref{fig9} presents comparisons of nonlinear simulations with linear theory at low density ratio $\rho_r=10^{-3}$ for zero-surface tension ($\alpha=0$), see section \ref{subsec:alphaZero}. A good agreement with linear theory is seen in fig. \ref{fig9} at various time instants up to steady-state, excepting a small region around $x=t$ where the IVP solution is divergent (see last paragraph of section \ref{subsubsec:asymp}). We note that these simulational results differ from \cite{Kataoka2024,Kataoka2026}, who study gravity waves forced by a wide hump and where (unlike our results) the authors find that the linear regime has very limited validity}.
     Fig. \ref{fig10} and \ref{fig11} provide comparisons with linear theory for low ($\rho_r = 10^{-3}$) and moderate density ratio ($\rho_r  = 10^{-1}$) respectively \textcolor{magenta}{for $\alpha >0$}. Very good match (\textcolor{magenta}{errors quantified below}) is seen in both figures in the spatial and temporal window of observation $ x \in \left[-5,10\right]$. In both figures, we observe that short waves upstream ($x<0$) are established \textcolor{magenta}{more rapidly} than the longer waves downstream, refer to panel (c) in both figures. Note that in the co-moving frame, the group velocities of $k_{l}$ and $k_s$ viz. $c_g^{(-)}(k_{s,l}) = \dfrac{1}{2} - \dfrac{\alpha k_{s,l}}{1 + \rho_r}$ (see Appendix \ref{AppB}) are nearly equal in magnitude but of opposite sign. Notice that at $t >> 1$ (panel (h) and (f) in figs. \ref{fig10} and \ref{fig11} respectively), the amplitude of upstream and downstream waves are nearly the same, albeit with very different wavelengths. While for fig. \ref{fig10}, we are able to reach nearly the steady-state at $t \approx 110$ (see solid black curve in panel (h) from expression \ref{eqIVP-12}), in fig. \ref{fig11}, we were able to validate linear theory in fig \ref{fig11} up to $t \approx 18.37 > 1$. For this fig. \ref{fig11}, beyond this time reflections were generated from the \textcolor{magenta}{computational} boundaries and thus these \textcolor{magenta}{results} are not presented here. These waves presumably arise from the larger inertia of the upper fluid in this regime of moderately large density ratio ($\rho_r=10^{-1}$); this is also consistent with the general observation that fig. \ref{fig10} (at air-water density ratio) presents a better agreement with linear theory than fig. \ref{fig11} (one hundred times larger $\rho_r$). The effect of changing grid is seen in both figures (adaptive with $2^{13}$ or $2^{14}$), showing negligible sensitivity on this scale. We have purposely suppressed the grid resolution image in the last panel of both figures, to avoid clutter. \textcolor{magenta}{In order to quantify the difference between theoretical predictions and simulations, we 
     	use the $L_2(t)$ error defined as $E_2(t) = \left[ \frac{\int_{-5}^{10} \left|\eta_{\mathrm{th}}(x,t) - \eta_{\mathrm{sim}}(x,t)\right|^2\,\mathrm{d}x}{\int_{-5}^{10} \left|\eta_{\mathrm{sim}}(x,t)\right|^2\,\mathrm{d}x} \right]^{1/2}$
     	where $\eta_{\mathrm{th}}$ and $\eta_{\mathrm{sim}}$ denote the theoretical (linear) and numerical solutions, respectively. For $\rho_r=10^{-3}$ in fig. \ref{fig10}, following the initial transient, $L_2(t)$ varies in the approximate range $0.11$-$0.13$. Similarly for $\rho_r=10^{-1}$ in fig. \ref{fig11}, this is
     	approximately $0.29$-$0.39$.} To our knowledge, figs. \ref{fig9}, \ref{fig10} and \ref{fig11} present the first benchmarking of analytical results for the time-dependent evolution to steady-state in the super-critical, two-dimensional RKL problem.
     
  	\subsection{Nonlinear regime}\label{subsec:nonlin}
  	\textcolor{magenta}{Having validated the linearised IVP predictions against simulations, we turn to nonlinear observations seen in the latter, as the strength of forcing is progressively increased. Appendix \ref{AppF-Basilisk} describes the benchmarking of Basilisk in the nonlinear regime by comparing the temporal evolution of a nonlinear Stokes wave using Basilisk and an unsteady, conformal mapping approach of \cite{Shelton_Milewski_Trinh_2025} showing good agreement.}
  	
\begin{figure}
	\centering
	\subfloat[$t=0.37$]{\includegraphics[width=0.47\textwidth]{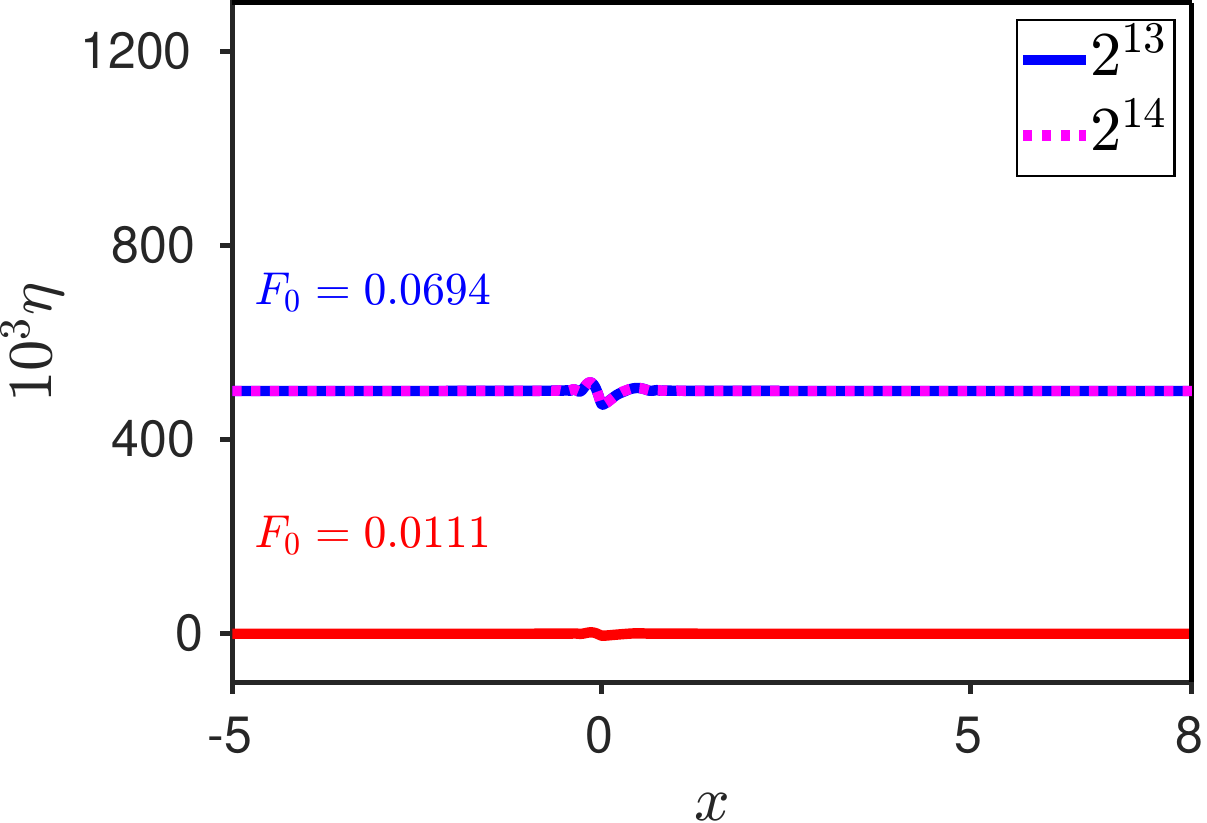} \label{fig12a}}
	\hspace{0.04\textwidth}
	\subfloat[$t=2.57$]{\includegraphics[width=0.47\textwidth]{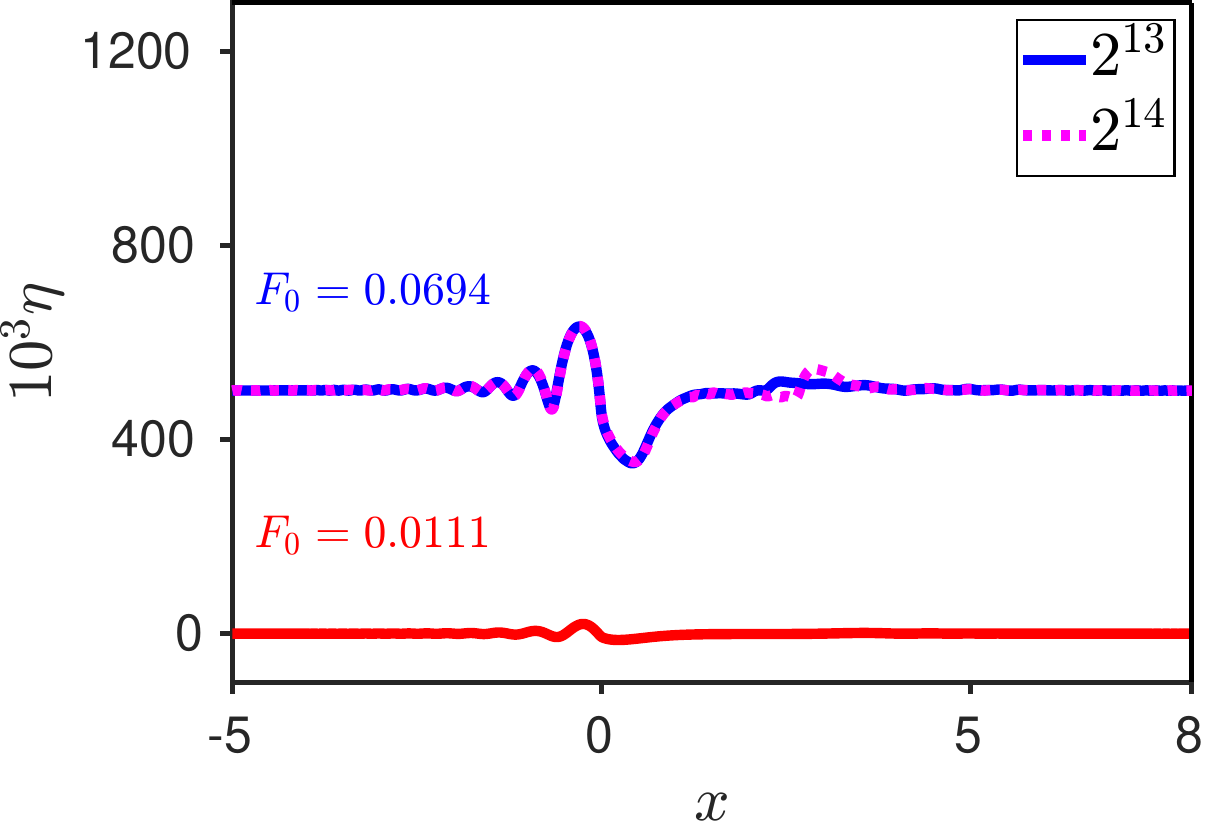} \label{fig12b}} \\
	\subfloat[$t=9.18$]{\includegraphics[width=0.47\textwidth]{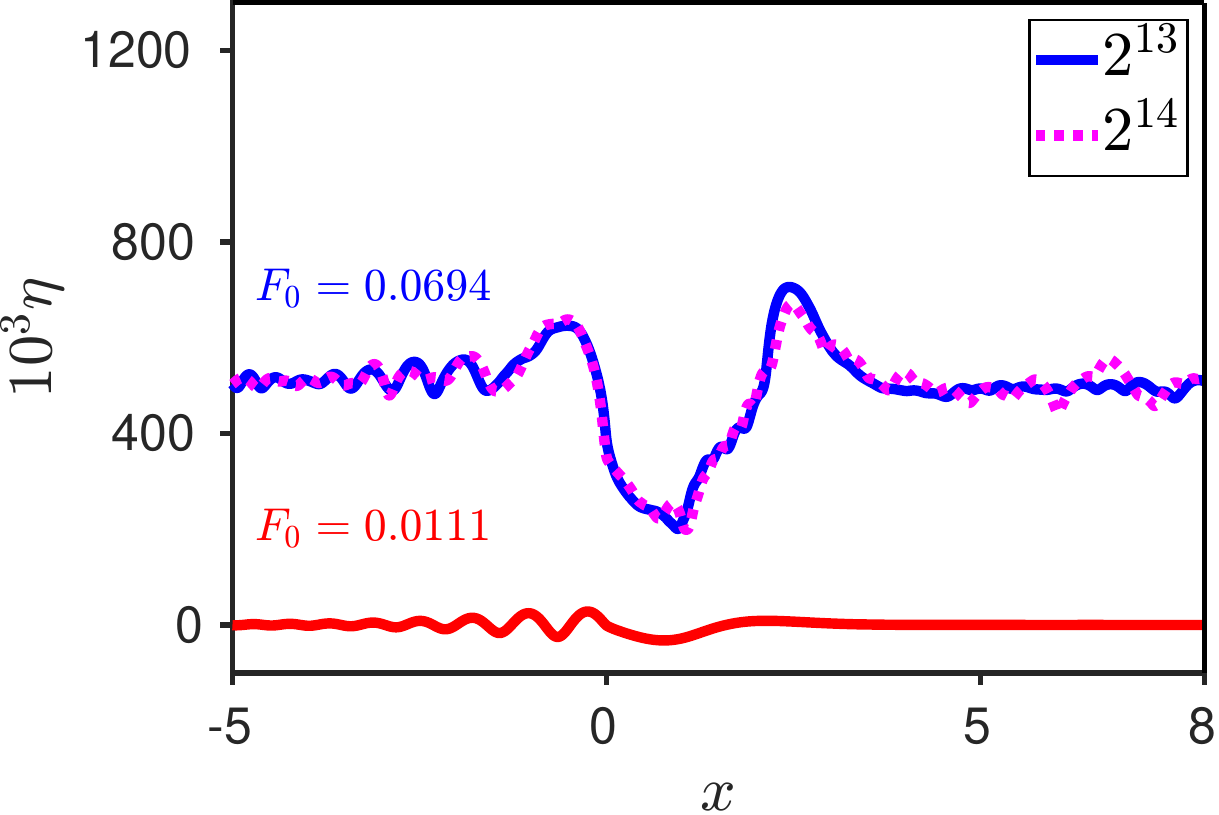} \label{fig12c}}
	\hspace{0.04\textwidth}
	\subfloat[$t=14.69$]{\includegraphics[width=0.47\textwidth]{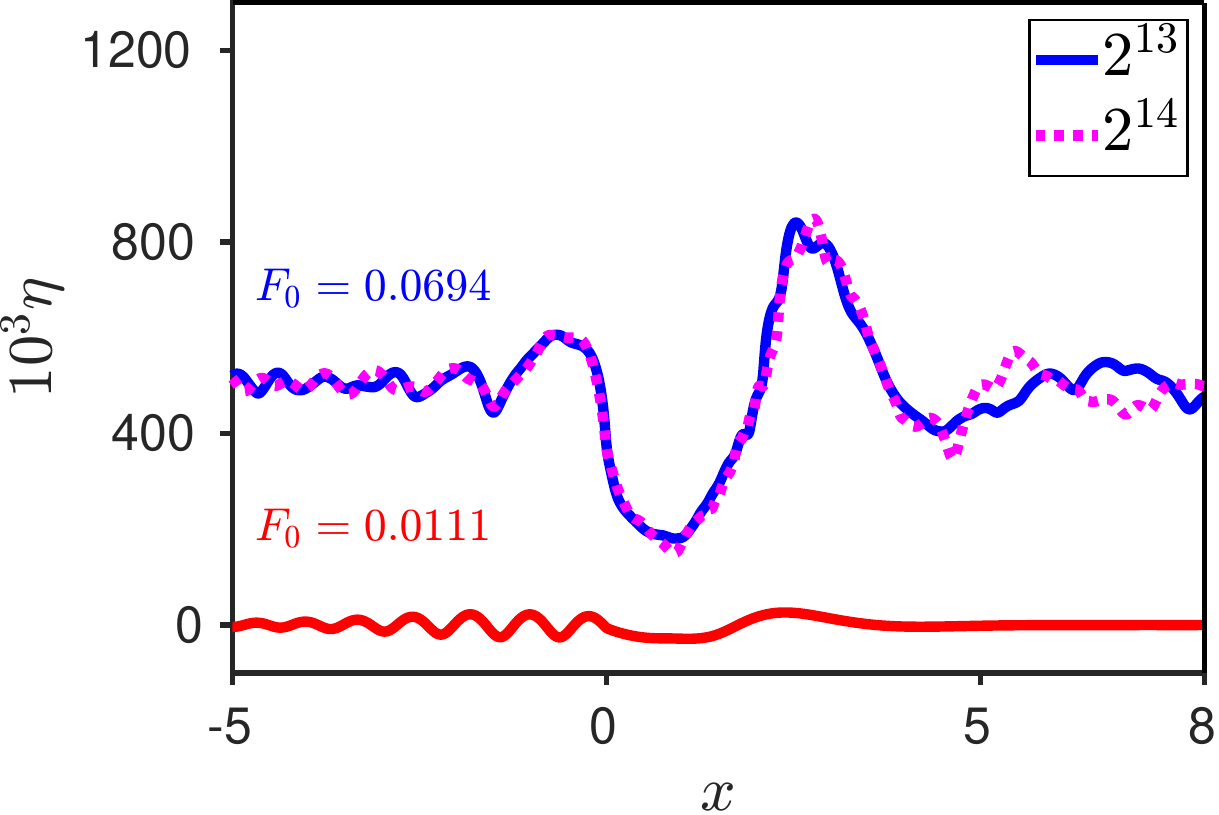} \label{fig12d}}\\
	\subfloat[$t=16.53$]{\includegraphics[width=0.47\textwidth]{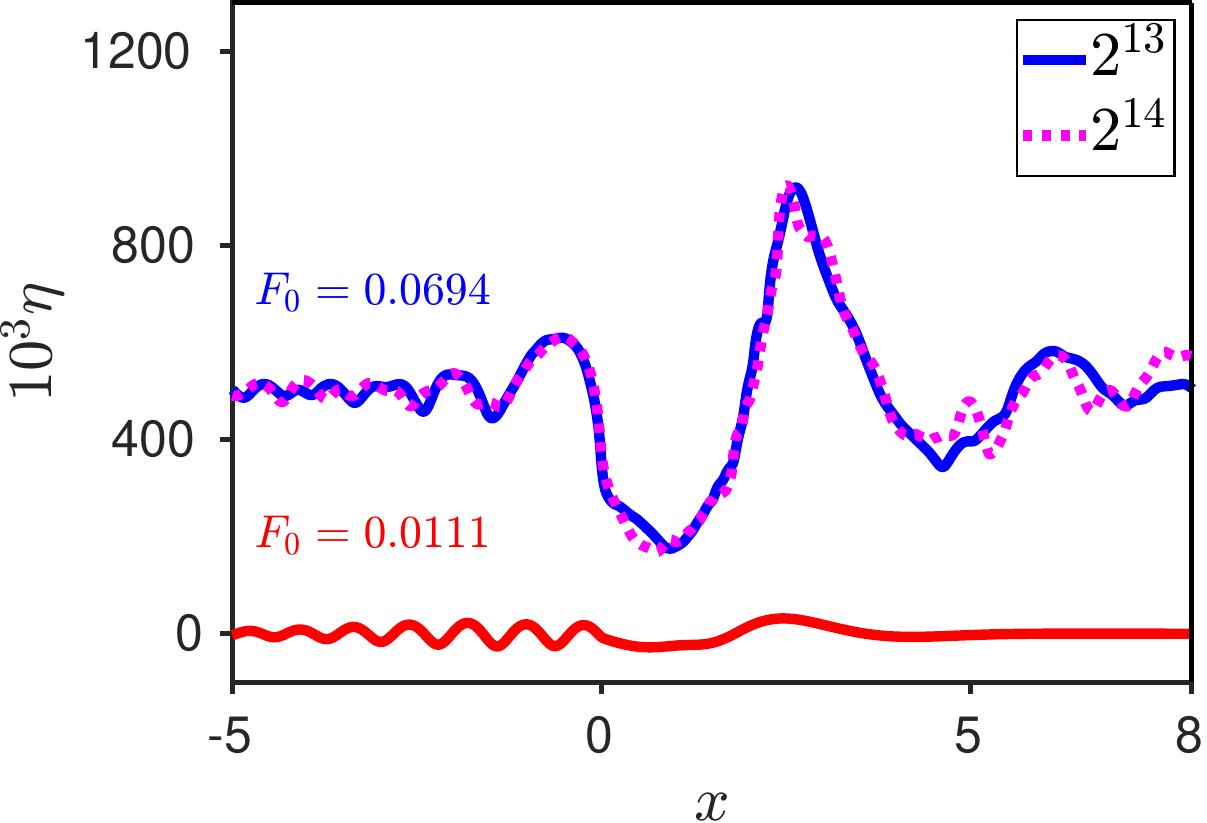} \label{fig12e}}
	\hspace{0.04\textwidth}
	\subfloat[$t=18.37$]{\includegraphics[width=0.47\textwidth]{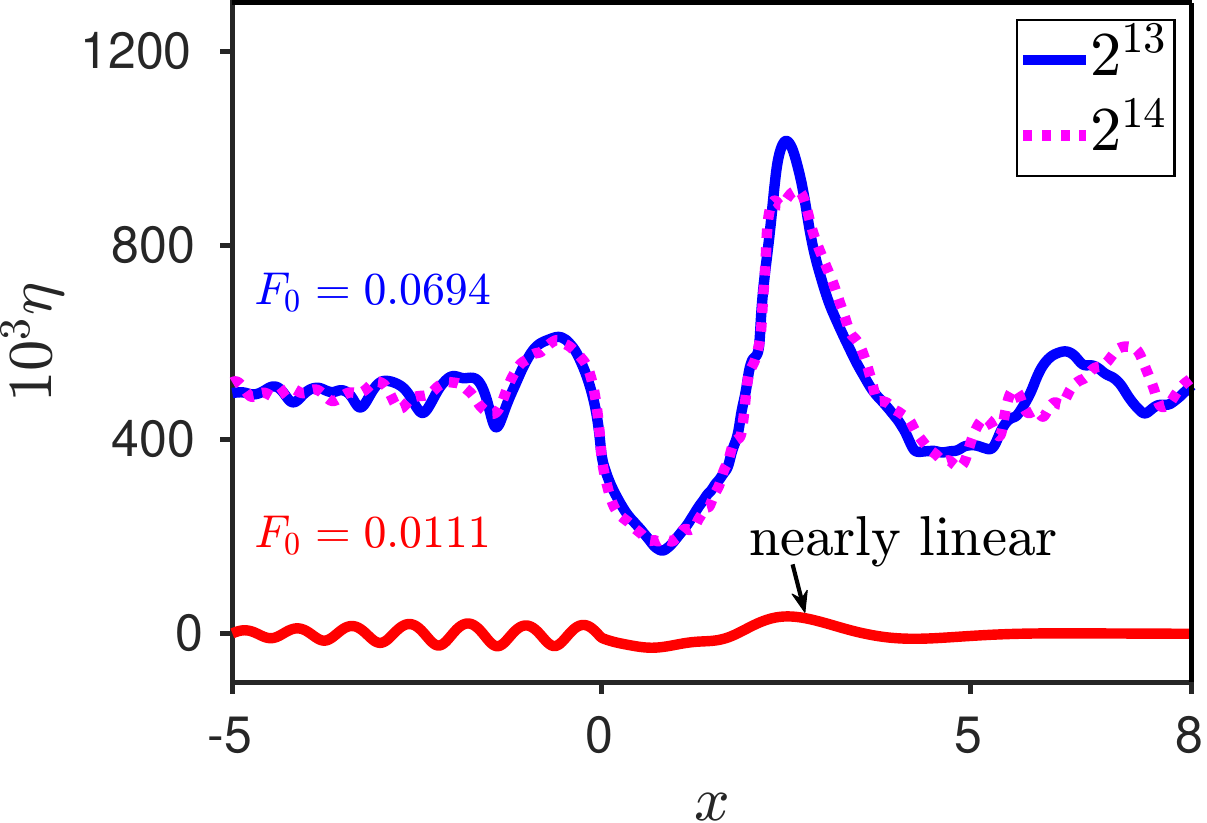} \label{fig12f}}\\      
	\captionsetup{justification=raggedright,singlelinecheck=false}
	\caption{\textcolor{magenta}{Time evolution $\rho_r=10^{-3},\alpha=0.1389$ and varying $F_0$. The second interface is shifted vertically upward by five hundred units. The lower curve(s) (red) show nearly linear behaviour.}}
	\label{fig12}
\end{figure}

  	Fig. \ref{fig12} presents the interface shape evolution for \textcolor{magenta}{two} choices of $F_0= \{1.1\times 10^{-2}, 6.94\times 10^{-2}\}$. The smaller value (curve in red), corresponds to nearly linear, temporal evolution discussed previously. Recall that in linear theory viz. expressions \ref{eqCan-5}(a)-(d), the strength of forcing $F_0$ is just a \textcolor{magenta}{scaling} factor. Fig. \ref{fig12}, shows that for $F_0 = 6.94\times 10^{-2}$, the shape of the interface differs qualitatively from the linear case (red curve with $F_0=1.1\times10^{-2}$). Evidently, the profiles with larger $F_0$ are \textit{not} scaled up versions of the \textcolor{magenta}{linear curves (in red). For these larger values of $F_0$, no steady-state is reached within the time window of observation. We note some grid dependency on the downstream side ($x>0$). In the region with longer waves ($x > 0$), note the sharp crests and rounded troughs in panels (e) and (f), and the opposite trends for $x<0$; these are reminiscent of Stokes and Crapper wave profiles \citep{Crapper1957} respectively. We refer the reader to similar sharp-crested profiles in fig. $1$ in \cite{schwartz1981nonlinear}, their results being at steady-state for the pure-gravity case.} 
  	  	
	 \begin{table}
		\centering  
		\setlength{\abovecaptionskip}{1pt}
		\setlength{\belowcaptionskip}{0pt}  
		\caption{Sectional summary of novel findings.}
		\label{tab2}
		\renewcommand{\arraystretch}{1.6}
		\small 
		\begin{tabularx}{\textwidth}
			{@{} >{\bfseries\color{black!80}}l X @{}}
			\toprule
			\textsf{Section \#} & \textsf{Key Result(s)} \\
			\midrule        
			
			\ref{sec:need} &
			\textcolor{magenta}{Utility for the time-dependent, IVP approach for wind-driven waves and other applications.} \\ 
			
			\ref{sec:IVP}, \ref{subsec:ivp_init}, \ref{subsec:steady}, Appendix \ref{AppC} &
			Analytical formulation of the IVP approach. Non-uniqueness of steady-state solutions. Extension of Lamb's formula to interfacial waves with $0 \leq \rho_r < 1$.\\ 
			
			\ref{sec:IVP-sol}, \ref{subsec:alphaZero} &
			Time-dependent expressions in the zero capillarity limit ($\alpha=0$),  their time-independent contribution and proof of upstream cancellation.\\ 
			
			\ref{subsubsec:asymp} &
			Algebraic decay of transient term(s) for $\alpha=0$. \\ 
			
			\ref{subsec:AlphGT0} & 
			IVP solutions for $\alpha >0$. Proof of up and downstream cancellations. Numerical Cauchy Principal Value of singular  integrals. \\ 
			
			\ref{sec:nonlin_sim} & 
			Comparison of linear theory with nonlinear simulations. Remarkable agreement in linear regime. \textcolor{magenta}{Qualitative observations of Stokes and Crapper-like waves in the nonlinear regime}. \\  
			
			Appendix \ref{AppD-GreenFn} \& \ref{AppE-Phillips} & 
			\textcolor{magenta}{Recovering the classical result of \cite{phillips1957generation} for random, point forcing in two-dimensions using the transient Green function and the IVP approach.}. \\  
			
			Online \citep{Kadari2026supplementary} & \textcolor{magenta}{Julia, MATLAB and Basilisk script files alongwith detailed documentation for asymptotic evaluation of Fourier integrals and their numerical evaluation.}
		\end{tabularx}  
		
	\end{table}

    \section{Conclusion}\label{sec:Concl}
    \textcolor{magenta}{In this study, we have analyzed the problem of forced interfacial waves using the linearised, IVP framework. This serves the purpose of elucidating the transient pathway towards steady-state solutions without having to invoke `\textit{ab extra}' conditions such as artificial viscosity or radiation boundary conditions. Furthermore, the analytical results of the transient approach including the method of Green's function, demonstrate their applicability for recovering the well-known results of \cite{phillips1957generation} for the case of stochastic pressure forcing.} Table \ref{tab2} summarises the key findings of our study. 

    Apart from fundamental interest, the time-dependent analytical expressions provided here at arbitrary density ratio $0 \leq \rho_r < 1$, are of potential utility for experimenters modelling wave phenomena at a density interface due to localised forcing e.g. the waves generated by a small boat at the salt-water, fresh-water interface have been used to model, for example, dead-water phenomena \citep{mercier2011resurrecting}. We provide the first benchmarking of linearised, analytical time-dependent predictions without employing radiation conditions or Rayleigh dissipation \textcolor{magenta}{in simulations}. Excellent agreement with linear theory, is seen when the forcing strength is small. This is reported at two density ratios, small ($10^{-3}$) and moderately large ($10^{-1}$). \textcolor{magenta}{A limited foray into the nonlinear regime has also been made at low density ratio ($\rho_r=10^{-3}$). Here, qualitative similarities between wave profiles observed in nonlinear simulations with Stokes and Crapper waves are highlighted.} 
    	
    \textcolor{magenta}{While we have considered the two-dimensional, transient problem here, three-dimensional effects are clearly important, particularly for the wind-wave problem. For example, \cite{caulliez1998generation} observed transient V-shaped surface wave streaks in their wind-wave experiments. The shape, strength of the pressure fluctuations and the corresponding speed of travel of Fourier components of the same (fig. $1$ in \cite{phillips1957generation}), can be estimated from coupled, wind-over-water simulations, see e.g. \cite{li2022principal}. Such three-dimensional simulations and their linearised, transient, analytical description are proposed as future work.}
        
    \noindent{\bf Funding\bf{:}} We gratefully acknowledge financial support from DST-SERB (Govt. of India) grants MTR/2019/001240, CRG/2020/003707, SPR/2021/000536, MoE-STARS/STARS-2/2023-0595 on various problems concerning nonlinear ocean waves, wave-breaking and spray formation. The doctoral tenures of VK (first author) and NY (second author) were supported by Ansys Inc. and the Prime Ministers Research Fellowship, Govt. of India and are sincerely acknowledged.\\
    
    \textcolor{magenta}{\noindent{\bf Usage of AI tools\bf{:}} We have used GPT-5.6 Sol \citep{openai2026gpt56sol} for assistance with the asymptotic expansion of Fourier integrals for $t\rightarrow\infty$, in section \S\ref{subsubsec:asymp}.}\\
    
    \noindent{\bf Declaration of Interests\bf{.}} The authors report no conflict of interest.

    \appendix 
	\section{Time-dependent equations}\label{AppA}
	The linearised, dimensional equations governing the temporal evolution of the disturbance velocity potential $\tilde{\vp}_l$ and $\tilde{\vp}_u$ and the perturbed interface $\tilde{\eta}$ are:
	\begin{subequations}\label{A1eq1}
		\begin{align}
			& \tilde{\nabla}^2\tilde{\vp}_u=0,\quad  -\infty < \tilde{x} < \infty,\quad \tilde{\eta}(\tilde{x},\tilde{t}) \leq \tilde{z} < \infty,  \tag{\theequation a}\\
			& \tilde{\nabla}^2\tilde{\vp}_l=0,\quad  -\infty < \tilde{x} < \infty,\quad -\infty < \tilde{z} \leq \tilde{\eta}(\tilde{x},\tilde{t}),  \tag{\theequation b}\\    	
			&\text{with boundary conditions (kinematic and dynamic):}\nonumber\\
			& \dfrac{\partial\tilde{\eta}}{\partial \tilde{t}}+ U\left(\dfrac{\partial\tilde{\eta}}{\partial \tilde{x}}\right) - \left(\dfrac{\partial\tilde{\vp}_u}{\partial \tilde{z}}\right)_{\tilde{z}=0} = \dfrac{\partial\tilde{\eta}}{\partial \tilde{t}}+ U\left(\dfrac{\partial\tilde{\eta}}{\partial \tilde{x}}\right) - \left(\dfrac{\partial\tilde{\vp}_l}{\partial \tilde{z}}\right)_{\tilde{z}=0} =  0,\quad \quad \tag{\theequation c,d} \\
			& - T\dfrac{\partial^2\tilde{\eta}}{\partial\tilde{x}^2} + \Bigg\{\rho_l\dfrac{\partial\tilde{\vp}_l}{\partial \tilde{t}} -  \rho_u\dfrac{\partial\tilde{\vp}_u}{\partial \tilde{t}} + U\rho_l\left(\dfrac{\partial\tilde{\vp}_l}{\partial \tilde{x}}\right) - U\rho_u\left(\dfrac{\partial\tilde{\vp}_u}{\partial \tilde{x}}\right)\bigg\}_{\tilde{z}=0} + \left(\rho_l - \rho_u\right)g\tilde{\eta} = -\tilde{p}_e(\tilde{x},\tilde{z}=0^{+},t), \tag{\theequation e} \\
			& \bm{\tilde{\nabla}}\tilde{\vp}_l(\tilde{z}\rightarrow -\infty,\tilde{t})\rightarrow \text{finite}\;,\bm{\tilde{\nabla}}\tilde{\vp}_u(\tilde{z}\rightarrow \infty,\tilde{t})\rightarrow \text{finite}. \tag{\theequation f}
		\end{align}
	\end{subequations}
	Here $\bm{\tilde{\nabla}} \equiv \left(\dfrac{\partial}{\partial \tilde{x}},\dfrac{\partial}{\partial \tilde{z}}\right)$ and the Laplacian operator is $\tilde{\nabla}^2 \equiv \dfrac{\partial^2}{\partial \tilde{x}^2} + \dfrac{\partial^2}{\partial \tilde{z}^2}$. Equations \ref{A1eq1} are first-order in time and require initial conditions which are chosen to be:
	\begin{subequations}\label{A1eq2}
		\begin{align}
			\tilde{\vp}_l(\tilde{x},\tilde{z},\tilde{t}=0)=\tilde{\vp}_u(\tilde{x},\tilde{z},\tilde{t}=0)=0,\; \tilde{\eta}(\tilde{x},\tilde{t}=0)=0. \tag{\theequation a,b,c}
		\end{align}
	We non-dimensionalise eqns. \ref{A1eq1} and \ref{A1eq2} using the length scale $l_c = U^2/g$, time scale $t_c=U/g$ and the pressure scale $p_c=\rho_lU^2$. Defining non-dimensional variables as follows,
	\begin{eqnarray}\label{A1eq3}
		\left(x,z,\eta\right) \equiv \dfrac{1}{l_c}		\left(\tilde{x},\tilde{z},\tilde{\eta}\right),\;\; t \equiv \dfrac{\tilde{t}}{t_c},\;\;p \equiv  \dfrac{\tilde{p}}{\rho_lU^2},\;\; \vp \equiv \dfrac{\tilde{\vp}\;t_c}{l_c^2},\;\; \bm{\nabla} \equiv l_c\bm{\tilde{\nabla}},
	\end{eqnarray}
	\end{subequations}
    we obtain the set of non-dimensional equations in \ref{eqIVP-1}.
    \section{Dispersion relation for unforced waves in the co-moving frame}\label{AppB}
\begin{figure}
     	\centering
     	\subfloat[$\rho_r = 10^{-3}$]{\includegraphics[scale=0.78]{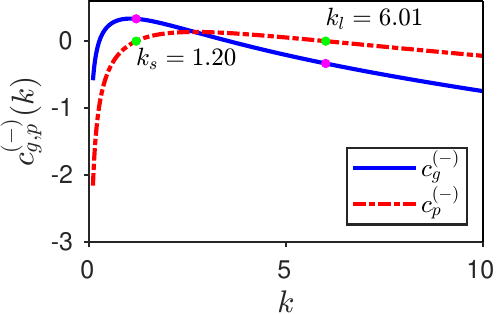}}\quad
     	\subfloat[$\rho_r = 10^{-1}$]{\includegraphics[scale=0.78]{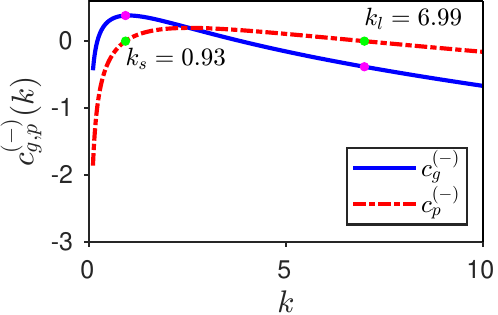}}
     	\caption{Phase ($c_p^{(-)}$) and energy propagation speed (group speed) $c_g^{(-)}$ for interfacial waves in deep layers from eqns. \ref{AppB-1} and \ref{AppB-2}. For both panels $\alpha=0.1389$. In the co-moving frame, the wavenumbers $k_{s,l}$ have zero phase speeds and positive and negative $c_g^{(-)}$ of same magnitude. Note that the gravity waves ($k_s$) become longer while capillary ones ($k_l$) get shorter with increase in density ratio.}
     	\label{fig13}
     \end{figure}

    Here we show that in the co-moving frame, Fourier modes $k_s$ and $k_l$ have energy propagation velocity which are positive and negative respectively.    In the co-moving frame, both upper and lower fluids move with uniform velocity (see fig. \ref{fig2}). The dispersion relation for unforced travelling waves is obtained from eqns. \ref{eqIVP-1}(a)-(f) (with $p_e(x,z=0^{+},t)=0$) by looking for solutions $\propto\exp\left[i(k x - \omega t)\right]$ where $k$ and $\omega$ are both non-dimensional. This is found to be:
    \begin{eqnarray}
    	\omega_{\pm} = k \pm \sqrt{\left(\dfrac{1-\rho_r}{1+\rho_r}\right)k + \dfrac{\alpha k^3}{1 + \rho_r}}. \label{AppB-1}
    \end{eqnarray}
    We now seek zero frequency solutions which are Fourier modes which can remain stationary in the co-moving frame. Such modes correspond to the negative branch of eqn. \ref{AppB-1} (negative sign) i.e. $\omega_{-}$. Setting $\omega_{-}$ equal to zero in \ref{AppB-1}, we find that these zero-frequency wavenumbers satisfy the quadratic equation $\alpha k^2 - \left(1+\rho_r\right)k + \left(1-\rho_r\right)=0$. This is the same quadratic equation found earlier. For $ 0 \leq \alpha < \alpha_{\text{max}}$, the roots $k_{s,l}$ to this quadratic are real (eqn. \ref{eqIVP-10}). The energy propagation velocity ($c_g^{(-)}$) for a Fourier mode with frequency $\omega_{-}$ may be shown to be:
    \begin{eqnarray}
    	c_g^{(-)}(k) \equiv \dfrac{d\omega_{-}}{dk} = 1 - \dfrac{\left(1 + \rho_r\right)^{-1/2}k^{-1/2}}{2}\left[\dfrac{3\alpha k^2 + \left(1-\rho_r\right)}{\left(\alpha k^2 + 1-\rho_r\right)^{1/2}}\right].  \label{AppB-2}
    \end{eqnarray}
     After some algebra, we find $c_g^{(-)}(k_{s,l}) = \dfrac{1}{2} - \dfrac{\alpha k_{s,l}}{1 + \rho_r}$ and from eqn. \ref{eqIVP-10} using the property $\dfrac{\alpha k_{s,l}}{1 + \rho_r}\lessgtr \dfrac{1}{2}$, we find $c_g^{(-)}(k_{s,l}) \gtrless 0$. Thus, in the linearised approximation, energy for the short waves ($k_l$) travel against the flow $c_g^{(-)}(k_{l}) < 0$ (leftwards or upstream in fig. \ref{fig2}), while that of longer waves ($k_s$) travel in the same direction as the flow (rightwards or downstream in fig. \ref{fig2}). Fig. \ref{fig13}, panels (a) and (b) show this at $\rho_r=10^{-3}$ and $\rho_r=10^{-1}$ respectively. The dots on the figures indicate $k_{s,l}$.
	\section{Rayleigh-Lamb dissipation approach for an interfacial wave}\label{AppC}
\begin{figure}
	\centering
	\subfloat[$x>0$]{\includegraphics[scale=0.55]{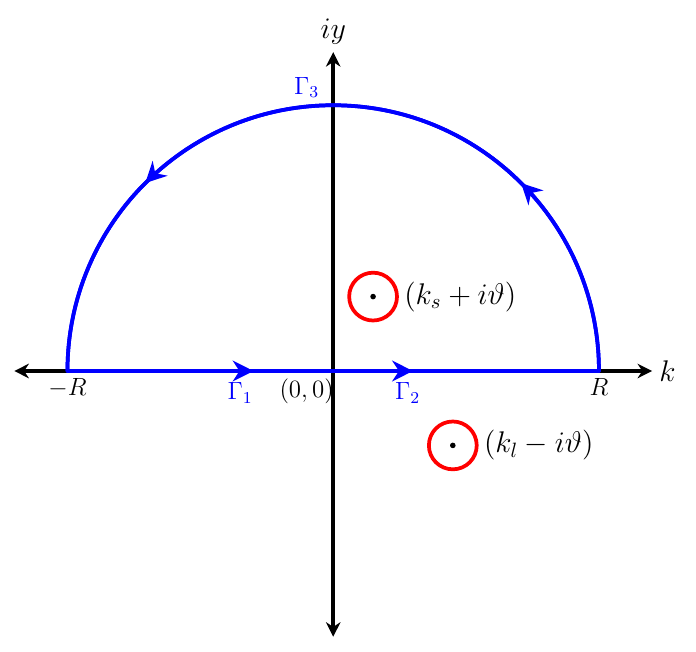}	\label{AppC_fig1}}
	\subfloat[$x<0$]{\includegraphics[scale=0.55]{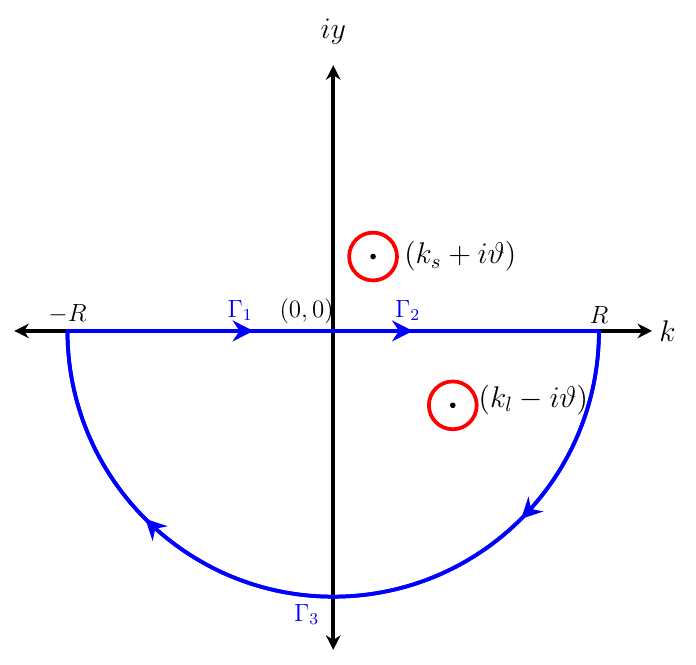}\label{AppC_fig2}}	
	\caption{Semicircular contour for (a) $x > 0$ (b) $x <0$ for evaluating the integrals in eqn. \ref{eq5.5}. The integrals on the semi-circular paths in both figures vanish as $R\rightarrow\infty$.}
	\label{fig14}
\end{figure}

	In this section, we prove that the steady-state expression in eqn. \ref{eqIVP-12} derived by \cite{lamb1932hydrodynamics} for $\rho_r=0$ (zero density ratio) remains unaltered for $0 \leq \rho_r < 1$, the only difference being that the roots $k_{s,l}$ get modified appropriately. The steady-state expression for $\bar{\eta}(k,t)$ is obtained by adding artificial `viscous' terms of the form $\mu \phi$ (with $\nu\rightarrow0$ at the end) to the Bernoulli equation \ref{A1eq1}e leading to
	\begin{eqnarray}\label{eq5.1}
		\bigg\{\gamma\left(\phi_{l} -\rho_r \phi_{u}\right) + \left(\dfrac{\partial\phi_{l}}{\partial x} -  \rho_r\dfrac{\partial\phi_{u}}{\partial x}\right)\bigg\}_{z=0} + \left(1-\rho_r\right)\eta - \alpha\left(\dfrac{\partial^2\eta}{\partial x^2}\right) = -p_e(x)
	\end{eqnarray}
	 where $\alpha \equiv \dfrac{gT}{\rho_l U^4},\;\gamma \equiv \dfrac{\mu U}{g}$ with $\gamma > 0$\footnote{invoking positive artifical viscosity, consistent with the principle of time-irreversibility and then setting the same to zero \citep{rayleigh1883form}, appears to be on the same logical footing as determining unknown constants by fitting to experimental data \citep{Kelvin1905}.}. The viscosity coefficient $\mu$ (distinct from dynamic viscosity) is chosen to be equal for both fluids and has the dimensions of $g/U$. This leads to:
	 \begin{eqnarray}
	 	\eta(x) &=& \dfrac{1}{\sqrt{2\pi}}\int_{-\infty}^{\infty}\; dk \exp\left(ikx\right)\dfrac{-\bar{p}_e(k)}{\alpha|k|^2 - \left(1+\rho_r\right)|k| + \left(1-\rho_r\right) + i\gamma(1+\rho_r)\dfrac{k}{|k|}} \nonumber \\
	 	&& +  \dfrac{1}{\sqrt{2\pi}}\int_{-\infty}^{\infty}\; dk \exp\left(ikx\right)\Bigg\{C_1 \delta(k-k_s^{*}) +  C_1^{*} \delta(k+k_s^{*})+ C_2 \delta(k-k_l^{*}) + C_2^{*}\delta(k+k_l^{*})\Bigg\},  \nonumber \\
	 	\label{eq5.2}
	 \end{eqnarray}
      where $k_s^{*}$ and $k_l^{*}$ are the slightly shifted counterparts of the real poles $k_{s,l}$, due to artificial dissipation $\gamma$, see fig. \ref{AppC_fig1} and \ref{AppC_fig2}. It is checked that in the limit $\gamma\rightarrow 0$, expression \ref{eq5.2} is equivalent to the first term on the right hand side of eqn. \ref{eqIVP-7} for $\bar{p}_e(k) = \dfrac{F_0}{\sqrt{2\pi}}$. As the poles do not lie on the path of integration, the integrals involving $C_1, C_1^{*}$ and $C_2, C_2^{*}$ evaluate to zero and we thus obtain,
     \begin{eqnarray}
     	\eta(x) &=& -\dfrac{F_0}{2\pi}\left\{\bigintsss_{0}^{\infty}\; dk \dfrac{\exp\left(ikx\right)}{\alpha\left(k-k_l\right)\left(k-k_s\right) + i\gamma(1+\rho_r)}\right. \nonumber \\
     	&&+ \left. \bigintsss_{0}^{\infty}\; dk \dfrac{\exp\left(-ikx\right)}{\alpha\left(k-k_l\right)\left(k-k_s\right) - i\gamma(1+\rho_r)}\right\},   \nonumber \\
     	&& \label{eq5.3} 	
     \end{eqnarray}
     where $k_{l,s} \equiv \dfrac{1+\rho_r}{2\alpha}\left(1 \pm \sqrt{1 - \dfrac{4\alpha\beta}{1 + \rho_r}}\;\right)$ with $\beta \equiv \dfrac{1-\rho_r}{1+\rho_r}$. Following \cite{lamb1932hydrodynamics}, the expression for $\eta(x)$ in \ref{eq5.3} may be re-written as
     \begin{eqnarray}
     	\eta(x) &=& \dfrac{F_0}{\alpha\pi}\int_{0}^{\infty}dk\;\dfrac{\left(k-k_s\right)\left(k_l-k\right)\cos(kx) - \Pi\sin(kx)}{\left(k-k_s\right)^2\left(k_l-k\right)^2 + \Pi^2}= \Re\left[\dfrac{F_0}{\alpha\pi}\int_{0}^{\infty}dk\;\dfrac{\exp(ikx)}{\left(k-k_s\right)\left(k_l-k\right)-i\Pi}\right]. \nonumber \\
     	&& \label{eq5.4}
     \end{eqnarray}
      where $\Pi \equiv \dfrac{\gamma\left(1+\rho_r\right)}{\alpha}$ and $\Re\left(\cdot\right)$ indicates the real part of its argument. Decomposing the integrand in eqn. \ref{eq5.4} into partial fractions, we obtain (neglecting a term of $\mathcal{O}(\tilde{\gamma}^2)$)
      \begin{eqnarray}
      	&&\eta(x) = \dfrac{F_0}{\pi \alpha\left(k_l-k_s - 2i\vartheta\right)}\Re\left[\int_{0}^{\infty}dk\exp(ikx)\biggl\{\dfrac{1}{k-(k_s + i\vartheta)} - \dfrac{1}{k-(k_l - i\vartheta)}\biggr\}\right]\nonumber \\
      	&&\label{eq5.5}
      \end{eqnarray}
  where $\vartheta \equiv \dfrac{\Pi}{k_l-k_s}$. For $x>0$, and using the contour shown in fig. \ref{AppC_fig1} on the complex $k$ plane, we evaluate the two integrals in eqn. \ref{eq5.5} using the Cauchy residue theorem, leading to
\begin{subequations}\label{eq5.6}
	\begin{align}
		&\int_{0}^{\infty} dk\frac{\exp(ikx)}{k-(k_{s}+i\vartheta)}=2\pi i\exp\bigg\{ix(k_{s}+i\vartheta)\bigg\}+\int_{0}^{\infty} dk\frac{\exp(-ikx)}{k+(k_{s}+i\vartheta)}, \nonumber \\
		&\text{and}\quad \int_{0}^{\infty} dk\frac{\exp(ikx)}{k-(k_{l}-i\vartheta)}=\int_{0}^{\infty} dk\frac{\exp(-ikx)}{k+(k_{l}-i\vartheta)}, \quad \left(x > 0,\vartheta>0\right). \tag{\theequation a,b}
	\end{align}
\end{subequations}
Similarly, for $x<0$ and using the contour shown in fig. \ref{AppC_fig2},we obtain
\begin{subequations}\label{eq5.7}
	\begin{align}
		&\int_{0}^{\infty}dk \frac{\exp\left(ikx\right)}{k-(k_{s}+i\vartheta)}=\int_{0}^{\infty}dk \frac{\exp\left(-ikx\right)}{k+(k_{s}+i\vartheta)}, 	\nonumber\\
		&\text{and}\quad \int_{0}^{\infty}dk \frac{\exp\left(ikx\right)}{k-(k_{l}-i\vartheta)}=-2\pi i\exp\bigg\{ix(k_{l}-i\vartheta)\bigg\}+\int_{0}^{\infty} dk\frac{\exp\left(-ikx\right)}{k+(k_{l}-i\vartheta)},\quad \left(x < 0,\vartheta>0\right).
\tag{\theequation a,b}
\end{align}
Taking the real part of the integrals in \ref{eq5.6} and \ref{eq5.7}, we obtain from expression \ref{eq5.5} with $\vartheta \rightarrow 0$, 
\end{subequations}
\begin{eqnarray}\label{eq5.8}
	\dfrac{\eta(x)}{F_0} = -\dfrac{2}{\alpha\left(k_{l} - k_{s}\right)}
	\left\{
	\begin{array}{lr}
		\sin(k_{l}x), & x<0\\
		\sin(k_{s}x), & x>0
	\end{array}
	\right\} + \dfrac{G(x)}{\pi\alpha},
\end{eqnarray}
where 
\begin{eqnarray}
	G(x) \equiv \dfrac{1}{k_{l}-k_{s}}\int_{0}^{\infty}\;dk\;\left\lbrace\dfrac{\cos(kx)}{k+k_{s}}-\dfrac{\cos(kx)}{k+k_{l}}\right\rbrace, \nonumber
\end{eqnarray}
and $k_{l,s}$ are (real) roots obtained from expressions in \ref{eqIVP-10}. 
{\color{magenta}
\section{Temporal Green function for eqn. \ref{eqIVP-5}}\label{AppD-GreenFn}
The (temporal) Green function for eqn. \ref{eqIVP-5} in the limit of $\rho_r = 0,\alpha=0$ with $P_0 = \dfrac{F_0}{\sqrt{2\pi}}$ is obtained as follows:
\begin{eqnarray}\label{D1}
\frac{\partial^2 A}{\partial t^2}
	& +\; 2i k \dfrac{\partial A}{\partial t}
	+ \bigg\{\alpha|k|^3 + \left(|k| - |k|^2\right)\bigg\} A
	= - i k\, \bar{p}_e(k,t) = -ik P_0\delta(t-t_0)
\end{eqnarray}
 With $A(k,t < t_0) = \dfrac{dA}{dt}(k,t< t_0) = 0$, it can be shown after some algebra that 
 \begin{eqnarray}\label{D2}
 	\bar{G}_A(k,t;t_0) \equiv A(k,t)  = -\dfrac{ikP_0}{|k|^{1/2}}\exp\left[-ik\left(t-t_0\right)\right]\sin\left[|k|^{1/2}\left(t-t_0\right)\right]H\left(t-t_0\right)
 \end{eqnarray}
   where $H(\cdot)$ is the Heaviside function. Knowing $\bar{G}_A(k,t;t_0)$, the corresponding Green function for the interface can be obtained from $\bar{G}_{\eta}(k,t;t_0) \equiv \bar{\eta}(k,t) = -\bar{p}_e(k,t) - \dfrac{\partial \bar{G}_A(k,t;t_0)}{\partial t} - ik\bar{G}_A(k,t)$ with
   \begin{eqnarray}\label{D3}
   	\bar{G}_{\eta}(k,t;t_0) = -P_0\delta(t-t_0) + ikP_0\exp\left[-ik(t-t_0)\right]\cos\left[|k|^{1/2}\left(t-t_0\right)\right]H(t-t_0)\nonumber \\
   \end{eqnarray}
   From linearity, we have 
   \begin{eqnarray}\label{D4}
   \bar{\eta}(k,t) = \int_{t_1}^{t}\bar{G}_{\eta}(k,t;t_0)\bar{p}_e(k,t_0)dt_0,\quad  t_0 < t
   \end{eqnarray}
    For steady, point forcing with $p_e(x,t) = F_0\delta(x)$ for all past time $t_0$ up to $t$ i.e. $\bar{p_e}(k,t_0)=\dfrac{F_0}{\sqrt{2\pi}}$, $-\infty < t_0 < t$. This implies that the lower limit of integration $t_1\rightarrow -\infty$ in  implying from \ref{D3} and \ref{D4}
    \begin{eqnarray}\label{D5}
    	\hat{\eta}(k,t) = &&P_0\int_{-\infty}^{t}\left[-\delta(t-t_0) + ik\exp\left[-ik\left(t-t_0\right)\right]\cos\bigg\{|k|^{1/2}\left(t-t_0\right)\bigg\}\right]dt_0
    \end{eqnarray}
    For fixed $t$, eqn. \ref{D5} can be transformed to a \textit{time-independent form} via change of variables $\tau \equiv t - t_0$ leading to 
    \begin{eqnarray}\label{D6}
    	\bar{\eta}(k) = P_0\left[-1 + \dfrac{ik}{2}\int_{0}^{\infty}\bigg\{\exp\left[-i\left(k-|k|^{1/2}\right)\right] + \exp\left[-i\left(k+|k|^{1/2}\right)\right]\bigg\}d\tau\right]
    \end{eqnarray}
Using the Abel regularised result $\int_{0}^{\infty}\exp\left[-iq\tau\right]d\tau = \pi\delta(q) - i PV\left(\dfrac{1}{q}\right)$ alongwith $\delta(k \mp |k|^{1/2}) = \delta(k\mp 1) + \delta(k)$, we obtain
\begin{eqnarray}\label{D7}
	\bar{\eta}(k) = P_0 \left[-1 + i\pi\bigg\{k\delta(k-1) + k\delta(k+1) + k\delta(k) - \dfrac{i}{\pi}PV\left(k\dfrac{\operatorname{sgn}(k)}{|k|-1}\right)\bigg\}\right]
\end{eqnarray}
With $k\delta(k)=0$ and $k\operatorname{sgn}(k)=|k|$, we have 
\begin{eqnarray}\label{D8}
	\bar{\eta}(k) = P_0 \left[ik\pi \bigg\{\delta\left(k-1\right) + \delta\left(k+1\right)\bigg\}+ PV\left(\dfrac{1}{|k|-1}\right)\right]
\end{eqnarray}
Recalling that $P_0 = \dfrac{F_0}{\sqrt{2\pi}}$, we note the structural similarity of eqn. \ref{D8} with the first of eqns. \ref{eqIVP-9}, described earlier for $\alpha=0,\rho_r=0$ and with $C_1=C_2=ik\pi$. It may be shown that \ref{D8}, upon Fourier inversion leads to 
\begin{eqnarray}\label{D9}
	\dfrac{\eta(x)}{F_0} = \left[-\sin(x) - \sin|x| + \dfrac{1}{\pi}\int_{0}^{\infty}\exp\left(-y|x|\right)\dfrac{y}{1+ y^2}dy\right], \quad -\infty < x < \infty
\end{eqnarray}
This result in eqn. \ref{D9} was also obtained earlier in section \ref{subsec:alphaZero} without usage of the Green function, as has been done here - see eqn. \ref{eqCan-3} and the term $\mathbb{T}_1(x)$ below this. 
\section{Recovering the \cite{phillips1957generation} result for random point forcing in two dimensions}\label{AppE-Phillips}
From eqn. \ref{D3}, the Green function (response at time $t$ due to a localised pressure impulse applied at $t_0 < t$) of eqn. \ref{eqIVP-5} for the interface, has the Fourier-space form (with $P_0=1$) (dropping the first term on the right hand side of \ref{D3}, for pulses received till time $t^{-}$) :
\begin{eqnarray}\label{E1}
	\bar{G}_{\eta}(k,t;t_0) = ik\exp\left[-ik(t-t_0)\right]\cos\left[|k|^{1/2}\left(t-t_0\right)\right], \quad t_0 < t
\end{eqnarray}
It may be noted from \ref{E1} that a non-zero interface response exists for any finite time, even on the negative side of the forcing, which rapidly decays to zero as $t\rightarrow\infty$. The Green function particularly has the advantage of estimating the cumulative effects of intermittent or time-continuous pressure forcing on the interface amplitudes on both sides. An excellent illustration of this is provided by considering the celebrated example of \cite{phillips1957generation} - of a stochastic pressure forcing and the resultant build-up of amplitude on a long time-scale. Towards this, consider the following argument. For time-dependent pressure impulses, we have from $\ref{D4}$ with $t_1=0$
\begin{eqnarray}\label{E2}
	\bar{\eta}(k,t) = \int_{0}^{t}\bar{G}_{\eta}(k,t;t_0)\bar{p}_e(k,t_0)dt_0
\end{eqnarray}
We now take $\bar{p}_e(k,t_0)$ to be Gaussian white noise with zero mean implying (long overbar indicates ensemble average and $^{*}$ indicates conjugation)
\begin{eqnarray}\label{E3}
	\overline{\bar{p}_e(k,t_0)}=0, \quad \overline{\bar{p}_e(k,t_0)\bar{p}^{*}_e(k,t_0^{'})} = \Gamma_k\delta(t_0-t_0^{'})
\end{eqnarray}
Defining a mean-squared amplitude,
\begin{eqnarray}\label{E4}
\overline{|\bar{\eta}(k,t)|^2}\equiv \overline{\bar{\eta}(k,t)\bar{\eta}^{*}(k,t)} = \int_{0}^{t}\int_{0}^{t}\overline{\bar{p}_e(k,t_0)\bar{p}^{*}_e(k,t_0)}\;\bar{G}_{\eta}(k,t;t_0)\;\bar{G}_{\eta}^{*}(k,t;t_0^{'})\;dt_0\;dt_0^{'} \nonumber \\
\end{eqnarray}
Using eqn. \ref{E3} in \ref{E4}, we obtain
\begin{eqnarray}\label{E5}
	\overline{|\bar{\eta}(k,t)|^2} = \Gamma_k \int_{0}^{t} \bar{G}_{\eta}(k,t;t_0)\;\bar{G}_{\eta}^{*}(k,t;t_0^{'}) dt_0 = \dfrac{\Gamma_k k^2}{2}\int_{0}^{t}\bigg\{1 + \cos\left[2\sqrt{|k|}(t-t_0)\right]dt_0\bigg\} \nonumber\\
\end{eqnarray}
Eqn. \ref{E5} may now be re-written after integration as
\begin{eqnarray}\label{E6}
	\overline{|\tilde{\eta}_k|^2} = \dfrac{\Gamma_kk^2}{2}t\;\bigg\{1 + \dfrac{1}{t}\dfrac{\sin\left(2\sqrt{|k|}t\right)-1}{2\sqrt{|k|}}\bigg\}
\end{eqnarray}
As $t\rightarrow\infty$, eqn. \ref{E6} shows that $\overline{|\bar{\eta}(k,t)|^2}\rightarrow \dfrac{\Gamma_kk^2}{2}t$, which is the \cite{phillips1957generation} result (see eqn. $4.13$ in \cite{phillips1957generation}), now recovered for our point forcing, in two-dimensions. Considering that for transient pulses, surface waves exist on either sides of the forcing, corresponding contributions to the mean-squared amplitude will appear from both sides of the forcing. This is unlike when the pressure is a constant where at steady-state, there are waves on one side only (for $\alpha=0$).
\section{Stokes travelling wave - Basilisk validation}\label{AppF-Basilisk}
\begin{figure}
	\centering
	\includegraphics[scale=0.15]{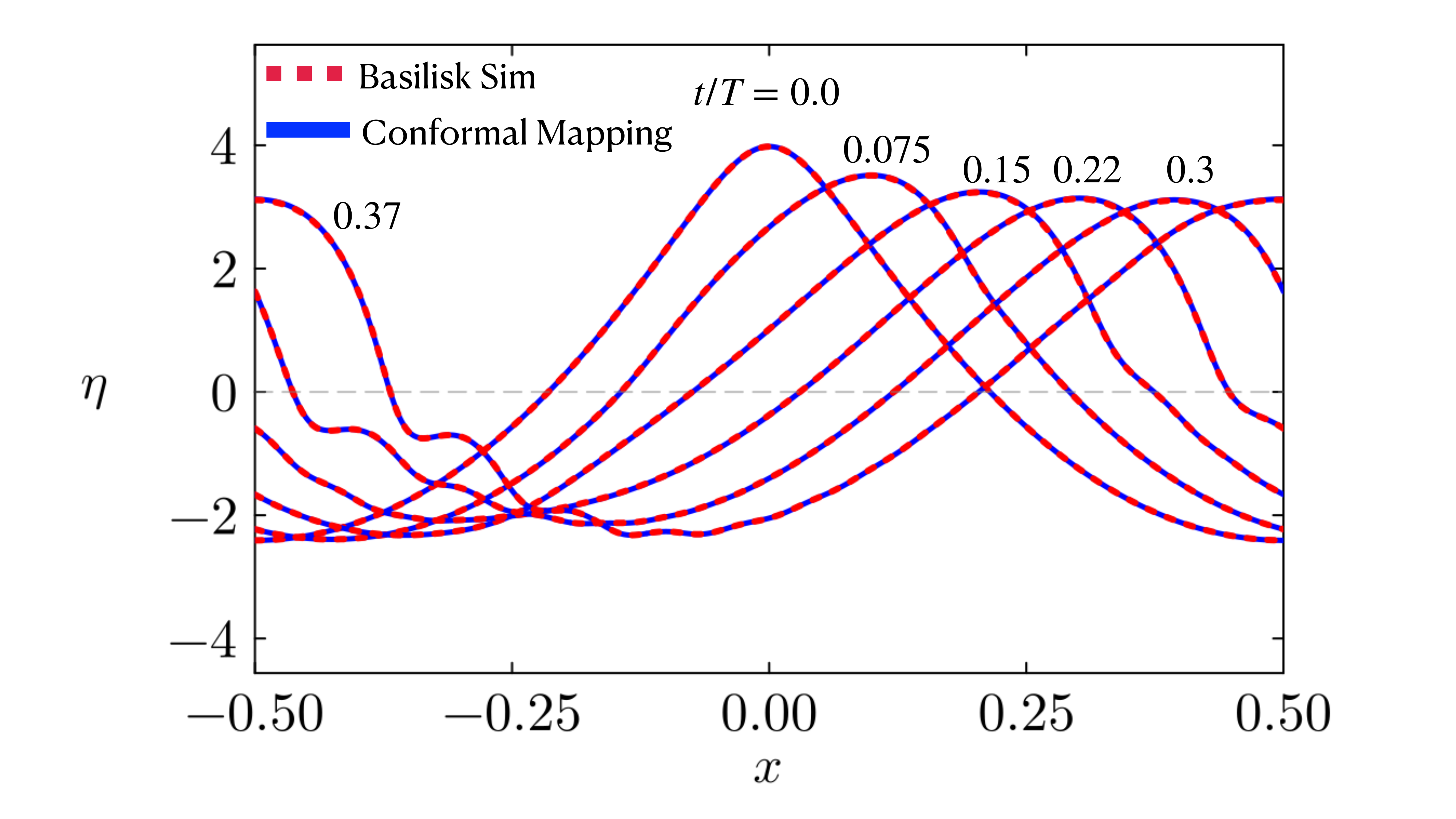}	\caption{\textcolor{magenta}{Comparison between time evolution of a Stokes wave via Basilisk and through unsteady, conformal mapping \citep{Shelton_Milewski_Trinh_2025}. The shape of the wave evolves in time, as capillarity has not been taken into account while computing its shape at $t=0$ (initial condition). Capillarity is however taken into account subsequently ($t>0$), in both Basilisk and the time-dependent conformal mapping approach. Note the excellent match between the wave shapes and the speed of travel at later time instants i.e $t/T=0.075,\;0.15,\;0.22,\;0.3,\;0.37$.}}
	\label{fig15}
\end{figure}
We show here that the two-phase solver Basilisk \citep{popinet2025basilisk} solving the incompressible Euler's equations with gravity and surface-tension, for air-water density ratio can describe accurately the temporal evolution of a nonlinear (Stokes) travelling-wave solution to the potential flow equations with a free-surface. Fig. \ref{fig15} shows a Stokes wave of steepness $0.38$ (product of amplitude and wavenumber) computed initially at $t/T=0$, where $T$ is the time taken by the wave to travel one wavelength. 

The initial shape of the Stokes wave and the associated velocity-field in the liquid below were computed using the steady (in the frame co-moving with the wave) potential flow equations $2.1$(a)-(d) in \cite{Shelton2021Structure}, while assuming surface-tension to be zero. This was done using the conformal mapping framework and the numerical method detailed in sections $2.1$ and $4$ of \cite{Shelton2021Structure} respectively, with zero surface-tension. The wave shape and the associated velocity field in the liquid (water), were seeded in Basilisk at time $t=0$ (initial condition). No velocities were provided on the air side in Basilisk at $t=0$. Air-water surface-tension was accounted for and the wave was propagated in time, by solving the incompressible Euler's equation with gravity and surface-tension in Basilisk. The same initial condition was also temporally evolved, via the time-dependent conformal mapping approach for the free-surface and the surface velocity potential. This was carried out numerically by following eqns. $2.4$(a)-(d) in \cite{Shelton_Milewski_Trinh_2025}. This numerical method for time evolution was as per appendix $A.1$ also in \cite{Shelton_Milewski_Trinh_2025}, while suppresssing viscous dissipation and wind-forcing in their formulation. For implementing these algorithms, codes were developed inhouse in Julia and are available at \citep{Kadari2026supplementary} (see section titled `Nonlinear-regime: Capillary-gravity wave evolution' in \cite{Kadari2026supplementary}). Fig. \ref{fig15} presents comparisons between the two time evolutions (Basilisk and the unsteady, conformal mapping approach) showing excellent agreement in both shape and speed.
}

	\bibliographystyle{jfm}%
	\bibliography{jfm}
\end{document}